\documentclass[pre,onecolumn,preprintnumbers,showpacs,floatfix,amsmath,amssymb]{revtex4}
\usepackage[utf8]{inputenc}
\usepackage{graphicx}
\usepackage{natbib}
\usepackage{subfigure}
\usepackage{color}

\begin{document}

\title{Thin film instability with thermal noise}
\author{ Javier A. Diez, Alejandro G. Gonz\'alez}
\affiliation{Instituto de Física Arroyo Seco (CIFICEN-CONICET), Universidad Nacional del Centro de la Provicia de Buenos Aires, Pinto 399, 7000, Tandil, Argentina}
\author{Roberto Fern\'andez}
\affiliation{Department of Mathematics, Utrecht University, P.O. Box 80010 3508 TA Utrech}

\begin{abstract}
We study the effects of stochastic thermal fluctuations on the instability of the free surface of a flat liquid film upon a solid substrate. These fluctuations are represented as a standard Brownian motion that can be added to the deterministic equation for the film thickness within the lubrication approximation. Here, we consider that while the noise term is white in time, it is coloured in space. This allows for the introduction of a finite correlation length in the description of the randomized intermolecular interaction. Together with the expected spatial periodicity of the flow, we find a dimensionless parameter, $\beta$, that accounts for the relative importance of the spatial correlation. We perform here the linear stability analysis (LSA) of the film under the influence of both terms, and find the corresponding power spectra for the amplitudes of the normal modes of the instability. We compare this theoretical result with the numerical simulations of the complete non-linear problem, and find a good agreement for early times. For late times, we find that the stochastic LSA predictions on the dominant wavelength remains basically valid. We also use the theoretical spectra to fit experimental data from a nanometric melted copper film, and find the corresponding times of the evolution as well as the values of the parameter $\beta$.
\end{abstract}

\maketitle

\section{Introduction}
\label{sec:Introd}
A basic problem in the study of free surfaces instabilities is the breakup of a flat thin liquid film on a solid substrate. Up to now, the description based on the hydrodynamic and deterministic Navier--Stokes equations are proven to be valid even down to the nanometric scale~\cite{ODB97}. This has been accomplished by introducing the intermolecular interaction between the liquid and the substrate. However, it is known that at these scales the thermal agitation of molecules is relevant when describing the behavior of matter~\cite{mecke_jpcm05,mosseler_sci70,hanggi_rmp09}. Thus, it is still necessary to investigate what role can play the thermal fluctuations in the hydrodynamic description of these instabilities. The consequences derived from considering this other effect can be of interest when designing microfluidic devices or electronic components whose function relies on thin film properties. In particular, we are interested on the effects that thermal noise may cause on films laterally much larger (up to microns) than their thicknesses.   

The study of the effects of thermal noise in the hydrodynamical equations was first introduced phenomenologically many years ago by Landau~\cite{LandLif} and Uhlenbeck~\cite{fox-uhlenbeck_pof70}. This inclusion can be done from the deterministic Boltzmann equation by a long-wave approximation, which justifies its microscopic feature. These equations have been used in the study of turbulence in randomly stirred fluids~\cite{forster_prl76}, the onset of instabilities in Rayleigh--Benard convection~\cite{hohenberg_pra92} and Taylor--Couette flow~\cite{swift_physA94}. This subject is of interest nowadays because one of the issues to be considered in the discussions about the differences between Navier--Stokes equation and molecular dynamics simulations is the effect of thermally triggered fluctuations in the classical hydrodynamic continuum modeling.

In particular, the application to unstable polymeric thin films has been the object of several theoretical and experimental studies~\cite{seeman_jphys05,seeman_prl01,fetzer_prl07}. In this problem, stochasticity has been analyzed using several techniques (such as Minkowsky invariants~\cite{mantz_jsm08}) to contrast some theoretical predictions with experimental results~\cite{becker_nat03}, where stochasticity is mostly considered as a spatial white noise. On the other hand, the problem of unstable liquid metal films with thermal noise has not been object of such a thorough study. In this problem, the solid coating is melted by laser, and  this introduces aspects that require the consideration of new factors such as the spatial correlation. Since the deposition of energy is not strictly uniform throughout the illuminated spot, the liquid lifetimes of different regions are not the same. In this context, there is a mix of factors to be considered when looking at samples from different regions. In fact, the time evolution of the sample at a certain region, due to the corresponding liquid lifetime, is compounded with the possibility that the laser illumination induces thermal fluctuations that might not be the same for all regions. One of the aims of the present paper is to consider how different spatial correlations of the ensuing fluctuations could influence the final spectra of the unstable modes. 

In this work we study the thin film instability by using a stochastic version of the thin-film equation based on the lubrication approximation for incompressible hydrodynamic equations (see Section~\ref{sec:TF_stoch}). Then, in Section~\ref{sec:LSA} we perform the linear stability analysis of the thin film under the perturbation with normal modes, and in Section~\ref{sec:num} we solve numerically the stochastic thin-film equation, and compare the results with the linear solution obtained previously. A comparison of theoretical predictions with experimental Fourier spectra obtained from SEM images of the instability of a melted copper film is presented in Section~\ref{sec:exp}, and finally we summarize and discuss the results in  Section~\ref{sec:conclu}.

\section{Thin film equations with stochastic noise}
\label{sec:TF_stoch}
In the framework of the continuous mechanics, the thermal agitation of the film molecules modifies the surface forces which describe the interaction between the fluid inside a volume element and its surroundings. Thus, an additional term, $\cal S$, in the expression of the Newtonian stress tensor has to be considered in order to include somehow the effect of molecular thermal motion~\cite{grun_jsp06,mecke_jpcm05}. Within the lubrication approximation, the most relevant component of $\cal S$ is ${\cal S}_{iz}$, where $i$ can be either  $x$ or $y$ and indicates a direction parallel to the substrate while $z$ stands  for the normal one. Due to its randomness, ${\cal S}_{iz}$ has zero mean 
\begin{equation}
\langle {\cal S}_{iz} (\vec x,t) \rangle = 0, 
\end{equation}
and the correlator is given by
\begin{equation}
 \langle {\cal S}_{iz}(\vec x,t) \, {\cal S}_{jz} (\vec x',t') \rangle= 2  \mu \, k_B T \, F(\vec x-\vec x') \, \delta(t-t') \,\delta_{i,j},
 \label{eq:corr_S}
\end{equation}
where $i,j=x,y$, $\mu$ is the fluid viscosity, and $\vec x=(x,y)$. Here, $k_B$ and $T$ are the Boltzmann constant and fluid temperature, respectively. This property is a consequence of the fluctuation-dissipation theorem of the statistical mechanics, which relates the fluctuations of physical quantifies to the dissipative properties of the system. From a physical point of view, the hydrodynamical equations are only valid at a scale large as compared to the molecular scale. Strictly speaking, since thermodynamic equilibrium is characterized by a Gaussian velocity spatial distribution, uncorrelated noise is required. Therefore, the correlation function $F(\vec x-\vec x')$ in Eq.~(\ref{eq:corr_S}) must have a small width. In the same approximation, the pressure terms in the isotropic part of the stress for a film of local thickness $h(\vec x, t)$ are given, as usually, by the capillary pressure, $-\gamma \nabla^2 h$ (where $\gamma$ is the surface tension), and the disjoining-conjoining pressure (van der Waals force), $\Pi(h)$. Thus, the reduction of the Navier--Stokes equations under the lubrication approximation leads to~\cite{grun_jsp06}:
\begin{equation}
3 \mu \frac{\partial h}{\partial t} + {\vec \nabla }\cdot \left[ h^{3} 
{\vec \nabla} \left( \gamma {\nabla^2 h} + \Pi(h) \right) \right]  
- {\vec \nabla }\cdot \left[ \int_0^h (h-z) {\cal S}_{||z}(z) dz\right]=0,
\label{eq:h_S}
\end{equation}
where ${\cal S}_{||z}=({\cal S}_{xz},{\cal S}_{yz})$. Note that the new noise term in Eq.~(\ref{eq:h_S}) enters as a rather complicated integral. It sums up $z$-uncorrelated noise terms over the film thickness, but it has the advantage that it maintains the conservative form of the equation. Thus, we have now a random current which acts as another driving force. It can be shown~\cite{grun_jsp06} that the  Fokker-Planck equation from  Eq.~(\ref{eq:h_S}) leads to the same time evolution  of the thickness distribution function as that of the Langevin equation
\begin{equation}
3 \mu \frac{\partial h}{\partial t} + {\vec \nabla }\cdot \left[ h^{3} 
{\vec \nabla} \left( \gamma {\nabla^2 h} + \Pi(h) \right) \right]  
- {\vec \nabla }\cdot \left[ \sqrt {3 h^3} \vec \xi(\vec x,t) \right]=0,
\label{eq:h}
\end{equation}
with a single multiplicative conserved noise vector $\vec \xi(\vec x,t)$, where the noise amplitude, $\vec \xi(\vec x,t)$, satisfies~\cite{grun_jsp06,fox-uhlenbeck_pof70}
\begin{equation}
 \langle \vec \xi(\vec x,t) \rangle=0, \qquad 
 \langle \xi_i(\vec x,t) \, \xi_j (\vec x',t') \rangle= 2  \mu \, k_B T \, F(\vec x-\vec x') \, \delta(t-t') \,\delta_{i,j}.
 \label{eq:corr_xi}
\end{equation}
In general, $\xi$ is a correlated noise in space, but a white noise in time.

Assuming symmetry along $y$-axis, the one dimensional version of Eq.~(\ref{eq:h}) for $h(x,t)$ is 
\begin{equation}
3 \mu \frac{\partial h}{\partial t} + 
\frac {\partial}{\partial x} \left[ h^{3} \left( \gamma \frac {\partial^3 h}{\partial x^3}+ 
\frac {\partial \Pi}{\partial x} \right) \right]  - 
  \frac {\partial}{\partial x} \left[ \sqrt{3 h^3} \xi(x,t) \right]=0,
\label{eq:h1_dim}
\end{equation}
where, for brevity, $\xi(x,t)$ stands for $\xi_x(x,t)$. 

Since the only characteristic length scale of an infinite film is its thickness, $h_0$, we define the following non dimensional variables,
\begin{equation}
 \tilde x= \frac {x}{h_0}, \quad \tilde y= \frac{h}{h_0}, \quad \tilde \xi = \frac {\xi}{\xi_0}, \quad 
 \tilde t= \frac{t}{t_0}, \quad \tilde \Pi = \frac{h_0}{\gamma} \Pi,
\end{equation}
where the scales of time, $t_0$, and noise, $\xi_0$, are to be determined in terms of the characteristic parameters of the problem. Note that we take the capillary pressure, $\gamma / h_0$, as the scale for the disjoining pressure. Thus, the dimensionless version of Eq.~(\ref{eq:h1_dim}) is:
\begin{equation}
\frac{\partial \tilde h}{\partial \tilde t} + 
\frac {\partial}{\partial \tilde x} \left[ {\tilde h}^{3} \left( \frac {\partial^3 \tilde h}{\partial {\tilde x}^3}+ 
\frac {\partial \tilde \Pi}{\partial \tilde x} \right) \right]  - 
\frac {\partial}{\partial \tilde x} \left[ \sqrt {3 {\tilde h}^3} \tilde \xi(x,t) \right]=0,
\label{eq:h1_ad}
\end{equation}
where
\begin{equation}
 t_0= \frac {3\mu h_0}{\gamma},
 \qquad
 \xi_0= \gamma h_0^{1/2}.
 \label{eq:t0_xi0}
\end{equation}

The temperature scale, $T_0$, can now be obtained from Eq.~(\ref{eq:corr_xi}). In fact, by defining the dimensionless temperature $\tilde T= T / T_0$, we obtain
\begin{equation}
 \langle \tilde \xi(x,t) \, \tilde \xi (x',t') \rangle= \tilde T \, F(x-x') \, \delta(t-t'),
 \label{eq:corr_xiad}
\end{equation}
with 
\begin{equation}
 T_0 = \frac {h_0^2 t_0 \xi_0^2}{2 k_B \mu}.
 \label{eq:T0}
\end{equation}

Moreover, it is still convenient to define the dimensionless noise amplitude as $\Theta = \tilde \xi / \sqrt{\tilde T}$, since the correlation 
\begin{equation}
 \langle \Theta(x,t) \, \Theta (x',t') \rangle= F(x-x') \, \delta(t-t')
 \label{eq:corr_Theta}
\end{equation}
is now normalized to one. Finally, the governing Eq.~(\ref{eq:h1_ad}) becomes
\begin{equation}
\frac{\partial \tilde h}{\partial \tilde t} + 
\frac {\partial}{\partial \tilde x} \left[ {\tilde h}^{3} \left( \frac {\partial^3 \tilde h}{\partial {\tilde x}^3}+ 
\frac {\partial \tilde \Pi }{\partial \tilde x} \right) \right]  - 
\sqrt {2 \sigma } \frac {\partial}{\partial \tilde x} \left[ {\tilde h}^{3/2} \Theta(\tilde x,\tilde t) \right]=0,
\label{eq:h1_ad1}
\end{equation}
where we define
\begin{equation}
 \sigma = \frac{3}{2} \tilde T = \frac {k_B T}{\gamma h_0^2},
\end{equation}
by using Eqs.~(\ref{eq:t0_xi0}) and (\ref{eq:T0}). As a result, we obtain a meaningful interpretation of the dimensionless constant $\sigma$. In fact, it gives the relative importance of the magnitude of the stochastic term (thermal noise) respect to the deterministic part of the equation is given by $\sigma$, in the form of the ratio between the thermal and surface energies of the system. Since typical experimental data yield $\sigma$ of the order of $10^{-4}$ (or even less) we will consider here this parameter within this range of values in order to look for effects on the film instability.

As regards to the form of $\Pi$, we take into account both the attractive and repulsive intermolecular liquid-solid forces, so that it includes both the disjoining and conjoining pressure terms in the form
\begin{equation}
 \Pi(h)=\kappa f(h)=\kappa \left[ \left( \frac {h_{\ast}}{h} \right)^3-\left( \frac {h_{\ast}}{h} 
\right)^2 \right],
\end{equation}
where $h_\ast$ is the dimensional equilibrium thickness, and $\kappa$ (with units of pressure) is given by
\begin{equation}
 \kappa = \frac{A}{6 \pi h_{\ast}^3}
\end{equation}
being $A$ the Hamaker constant. In dimensionless variables, $\kappa$ becomes $K = \kappa h_0 / \gamma$, and then the final version of Eq.~(\ref{eq:h1_ad1}) is:
\begin{equation}
\frac{\partial h}{\partial t} + 
\frac {\partial}{\partial x} \left[ {h}^{3} \left( \frac {\partial^3 h}{\partial {x}^3}+ 
K f'(h) \frac {\partial h}{\partial x} \right) \right]  - 
\sqrt {2 \sigma } \frac {\partial}{\partial x} \left[ h^{3/2} \Theta(x,t) \right]=0,
\label{eq:h1_ad2}
\end{equation}
where we omit the tilde (~$\tilde {}$~) for brevity and for now on.

For the stochastic term, we consider as usual that $\Theta (x,t)$ is related to a standard Brownian motion as
\begin{equation}
\Theta (x,t)\;=\; \frac{\partial W(x,t)}{\partial t}, 
\label{eq:Th_Wt}
\end{equation}
which satisfies 
\begin{equation}
W (x,t+\Delta) - W (x,t) \;\sim\;  {\cal N}(0,\Delta),
\label{eq:W_Delta}
\end{equation}
where ${\cal N}(0,\Delta)$ is a normal distribution with zero mean and \emph{variance} $\Delta$. Here, the notation ``$\sim$'' means an equality of distributions. 

\section{Linear stability analysis (LSA) of the stochastic thin film equation}
\label{sec:LSA}
At the beginning of the instability process the deviations, $\delta h(x,t) = h(x,t)- \tilde h_0$, from the initial average film height are small (even if $\tilde h_0=1$, we keep this notation for clarity). By expanding Eq.~(\ref{eq:h1_ad}) up to first order in $\delta h$ and $\Theta$ (assuming that the noise amplitude is small as well) we obtain the linear stochastic equation,
\begin{equation}
\frac{\partial \delta h}{\partial t} + 
\tilde h_0^{3}\left( \frac{\partial^4 \delta h}{\partial x^4}  + 
K f'(\tilde h_0) \frac{\partial^2 \delta h}{\partial x^2} \right)   
- \sqrt{2 \sigma {\tilde h_0}^3 } \, \frac{\partial \Theta}{\partial x} =0.
\label{eq:dhad}
\end{equation}
It is convenient to look for its solution in the Fourier space, so that we have
\begin{equation}
\frac {\partial \delta\widehat h(q,t)}{\partial t} = 
\omega(q) \widehat {\delta h}(q,t) + i \sqrt{2 \sigma {\tilde h_0}^3 } \, q \widehat \Theta
\label{eq:fou1}
\end{equation}
where the Fourier transform is defined by
\begin{equation}
\widehat {\delta h}(q, t)= \int_{-\infty}^{\infty} \delta h(x,t) \, e^{-i q x} \, dx,
\end{equation}
and $\omega(q)$ is the deterministic dispersion relation. This one is given by~\cite{dk_pof07}
\begin{equation}
 \omega(q)=4 \omega_m \left[ \left( \frac{q}{q_c}\right)^2- \left(\frac{q}{q_c}\right)^4 \right].
 \label{eq:w_det}
\end{equation}
where
\begin{equation}
 q_c=\sqrt{K f'(\tilde h_0)}, \quad \omega_m=\frac {{\tilde h_0}^3 q_c^4}{4}
 \label{eq:qc_wm}
\end{equation}
are the critical (marginal) wavenumber and the maximum growth rate, respectively. The wavenumber of maximum growth rate is $q_m=q_c/\sqrt{2}$.

Since Eq.~(\ref{eq:fou1}) is an equation of the Langevin type, its solution is given by~\cite{Chow_book,Evans},
\begin{equation}
\widehat {\delta h}(q,t)\;=\; e^{\omega(q)t} \, \widehat{\delta h}(q,0) + i\sqrt{2 \sigma {\tilde h_0}^3} q \int_0^t  e^{\omega(q)(t-s)} d\widehat W(q,s)\;.
\label{eq:sol1}
\end{equation}

In order to study the instability evolution in the spectral space, we calculate the autocorrelation
\begin{equation}
\langle\,\widehat {\delta h}(q,t)\,\widehat {\delta h}(q',t')\,\rangle \;=\; A_1 + A_2 + A_3 + A_4\,,
\label{eq:corr_hh}
\end{equation}
where the terms on the r.h.s. are defined as follows:
\begin{eqnarray}
\label{eq:A1234}
A_1&=& \langle\,\widehat {\delta h}(q,0)\,\widehat {\delta h}(q',0)\,\rangle \,e^{\omega(q)t}\,e^{\omega(q')t'},\\ \nonumber
A_2 &\propto& \bigl\langle \widehat{\delta h}(q,0)\, d\widehat W(q',t')\bigr\rangle, \\ \nonumber
A_3 &\propto & \bigl\langle \widehat {\delta h}(q',0)\, d\widehat W(q,t)\bigr\rangle, \\
A_4 &=&-2 \sigma \, {\tilde h_0}^3 \, q^2  
\langle\, \int_0^t  e^{\omega(q)(t-s)} d\widehat W(q,s)\, 
\int_0^{t'}  e^{\omega(q')(t'-s')} d\widehat W(q',s') \,\rangle\;.\nonumber
\end{eqnarray}

In order to calculate these terms, let us first consider the autocorrelation of the Fourier transformed noise, $\widehat \Theta$, as,
\begin{eqnarray}
\langle\, \widehat\Theta(q,t) \, \widehat\Theta(q',t') \,\rangle &=& 
\int_{-\infty}^{\infty} \int_{-\infty}^{\infty} \langle\, \Theta(x,t) \, \Theta(x',t') \,\rangle \,e^{-i q x}e^{-i q' x'} dx\,dx'
\nonumber\\
&=& \int_{-\infty}^{\infty} \int_{-\infty}^{\infty} \,\delta(t-t')\,F(x-x')\, e^{-i (q x+ q' x')}\, dx\,dx'\,\\
&=& \,2\pi \, \delta(q+q') \,\delta(t-t') \widehat F(q) \,\nonumber
\label{eq:xx}
\end{eqnarray}
where  
\begin{equation}
\widehat F(q) = \,\int_{-\infty}^{\infty} F(u) e^{-i q u} du\,
\label{eq:chi}
\end{equation}
is the Fourier transform of the correlation function, $F(u)$, being $u=x-x'$, and we have used 
$\int_{-\infty}^{\infty} e^{-i q x} dx = 2\pi \,\delta(q)$. 

Note that only in the case of non-correlated noise we have $\widehat F(q)=1$, otherwise this transform has to be calculated (see Section~\ref{sec:corr}). Since $\widehat\Theta$ is a white noise in time, its Fourier transform satisfies (see Eq.~(\ref{eq:Th_Wt}))
\[
\widehat\Theta(q,t)\;=\; \frac{\partial \widehat W(q,t)}{\partial t}.
\]
Then, the autocorrelation of $\widehat W(q,t)$ is given by
\begin{equation}
\langle\, \widehat W(q,t) \,\widehat W(q',t')\rangle \;=\; 2\pi \, \delta(q+q')\,\widehat F(q) (t\wedge t'),
\label{eq:WW}
\end{equation}
where $t \wedge t'$ stands for the minimum of $t$ and $t'$.

On the other hand, the height--height correlation for the initial condition is
\begin{eqnarray}
\langle\,\delta \widehat h(q,0)\,\delta \widehat h(q',0)\,\rangle &=& 
\int_{-\infty}^{\infty} \int_{-\infty}^{\infty} \langle\,\delta h(x,0)\,\delta h(x',0)\,\rangle \,e^{-i q x}e^{-i q' x'} dx\,dx'
\nonumber\\
&=&\int_{-\infty}^{\infty} \int_{-\infty}^{\infty}  F_0(u)\,e^{-i q u} \,  e^{-i (q+q') x'} dx'\, du\,
\nonumber\\
&=&  2\pi \, \widehat F_0(q)   \, \delta(q+q')  \;.
\label{eq:corr_hh0}
\end{eqnarray}
where $F_0(u)$ is the spatial correlation function of the initial condition, and $\widehat F_0(q)$ its Fourier transform.

Consequently, we can write $A_1$ in Eq.~(\ref{eq:A1234}) as
\begin{equation}
A_1 =  2\pi \, \widehat F_0(q) \, \delta(q+q') \,e^{\omega(q)(t+t')}\;,
\label{eq:A1}
\end{equation}
where we have considered the parity $\omega (-q)=\omega (q)$.

Regarding $A_2$ and $A_3$ in Eq.~(\ref{eq:A1234}), we note that 
\begin{equation}
 A_2=A_3=0
 \label{eq:A23}
\end{equation}
because the randomness of the initial condition is independent of the Brownian $W$. 

For the term $A_4$ in Eq.~(\ref{eq:A1234}) we note that since the Brownians in different time intervals $t$ and $t'$ are not correlated, only the common interval $[0,t \wedge t']$ contributes to the correlation of the product of the integrals. Besides, due to Eq.~(\ref{eq:WW}) only the terms with $q'=-q$ have non zero correlation. Thus, we obtain
\begin{eqnarray}
A_4 &=& -2 \sigma \, {\tilde h_0}^3 \,q^2\, 2\pi \,\delta(q+q')
E\biggl[ \int_0^{t\wedge t'}  e^{\omega(q)(t-s)} d\widehat W(s)\, 
\int_0^{t\wedge t'}  e^{\omega(q)(t'-s')} d\widehat W(s') \biggr] \nonumber \\
&=& -2 \sigma \, {\tilde h_0}^3 \,q^2\,\delta(q+q') \widehat F(q)
\int_0^{t\wedge t'}  e^{\omega(q)(t-s)}\, e^{\omega(q)(t'-s)} ds\;.
\label{eq:A4}
\end{eqnarray}
The last line above is consequence of one of the lemmas of Ito's integral~\cite{Chow_book,Evans}, since the stochastic process in time is a white noise. Performing the integral and using $t+t'-2(t\wedge t') = |t-t'|$, we have
\begin{equation}
A_4 \;=\; \sigma\,  {\tilde h_0}^3 \,2\pi \,\delta(q+q')\, \frac{ q^2 \widehat F(q)}{\omega(q)}\Bigl[
e^{\omega(q)(t+t')} - e^{\omega(q)|t-t'|}\Bigr]
\label{eq:A4_new}
\end{equation}
Finally, by replacing Eqs.~(\ref{eq:A1}), (\ref{eq:A23}) and (\ref{eq:A4_new}) into Eq.~(\ref{eq:corr_hh}), we obtain
\begin{equation}
\langle\,\delta \widehat h(q,t)\,\delta \widehat h'(q',t')\,\rangle =  2\pi \,\delta(q+q') S(q;t,t'),
\label{eq:corr_hh_new}
\end{equation}
where
\begin{equation}
 S(q;t,t')= \widehat F_0(q) \,e^{w(q)(t+t')}\, + 
 \,\sigma\, {\tilde h_0}^3 \, \frac{ q^2 \widehat F(q)}{\omega(q)}\Bigl[e^{\omega(q)(t+t')} - e^{\omega(q)|t-t'|}\Bigr].
 \label{eq:S}
\end{equation}
For the case of non-correlated noise, we have $\widehat F(q)=1$, which reduces this equation to that obtain in~\cite{mecke_jpcm05}. Note that the first term of Eq.~(\ref{eq:S}) corresponds to the spectra predicted by the deterministic model ($\sigma=0$). Here, we aim to compare the evolution of films with ($\sigma > 0$) and without ($\sigma=0$) the stochastic term. To study the corresponding spectra separately the film has to be perturbed at $t=0$, otherwise the film does not evolve in the deterministic case. Thus, we consider that the originally flat free surface of the film is slightly modified by a small amplitude perturbation of the form,
\begin{equation} 
 \delta h(x,0) = \sum_{k=1}^{N} B_k \sin (2\pi x k/L),
 \label{eq:delta0}
\end{equation}
and the amplitudes $B_k$ are random numbers with $|B_k|<B_{max}=10^{-3} \tilde h_0$. 

As a typical case, in the following calculations we choose a film with $h_\ast=0.1$ and $\theta=30^\circ$, which yields~\cite{dk_pof07} $q_m=0.151$, $q_c=0.213$ and $\omega_m=5.19~10^{-4}$. The quantities $\lambda_m= 2 \pi /q_m = 41.6$ and $\tau_m=(1/\omega_m) \ln [{(\tilde h_0-h_\ast)/B_{max}}]= 13113.5$ give a rough idea of the spatial extension and time duration of the film breakup process. We find that $L=500\approx 12\lambda_m$ is large enough to produce results which are practically domain size independent. The consequences on the stochastic process of using a correlated noise on a finite domain is analyzed in the next section.

\subsection{Correlated stochastic noise in a finite domain}
\label{sec:corr}

Here, we will assume that the correlation function $F(x-x')$ in Eq.~(\ref{eq:corr_xi}) is $L$-periodic. Moreover, we give meaning to the stochastic process $\Theta(x,t)$ in terms of a Q--Wiener process in the form (see Eq.~(\ref{eq:Th_Wt}))
\begin{equation}
 \Theta(x,t)=\frac {\partial W(x,t)}{\partial t}= \sum_{k=-\infty}^{+\infty} \chi_k \dot \beta_k(t) g_k(x),
 \label{eq:zeta}
\end{equation}
where $\beta_k$ (with $k$ integer) forms a family of mutually independent Brownian motions with respect to time and the dot stands for time derivative. The constants $\chi_k$ are the eigenvalues of the Hilbert-Schmidt operator ${\cal Q}$, defined by:
\begin{equation}
 {\cal Q} f(x) = \int_{-L/2}^{L/2} F(x-x') f(x') dx'.
\end{equation}
The corresponding complete system of orthonormal eigenfunctions, $g_k(x)$, that satisfy 
\begin{equation}
 {\cal Q} g_k(x) = \chi_k g_k(x)
 \label{eq:Qg}
\end{equation}
are 
\begin{equation}
 g_k(x)=\left\{
 \begin{tabular}{lr}
  $\sqrt{\frac {2}{L}} \cos \left( q_ k x \right),$ & $k>0$ \\
  $\sqrt{\frac {1}{L}},$ & $k=0$ \\
  $\sqrt{\frac {2}{L}} \sin \left( q_ k  x\right),$ & $k<0$ \\
 \end{tabular}
  \right.
  \label{eq:gk}
\end{equation}
In fact, this can be easily verified that by considering the complex eigenfunction $G_k(x)=e^{-i q_k x}$ with $q_k=2 \pi k / L$. By defining $u=x-x'$, we obtain
\begin{equation}
{\cal Q} G_k(x)=\int_{-L/2}^{L/2} F(x-x') e^{-i q_k x'} dx'= e^{-i q_k x} \int_{-L/2-x}^{L/2-x} F(u) e^{-i q_k u} du=\chi_k G_k(x),
\end{equation}
with the eigenvalue given by
\begin{equation}
 \chi_k=\int_{-L/2}^{L/2} F(u) e^{-i q_k u} du.
 \label{eq:sigmak0}
\end{equation} 
Here, we used the parity property $F(u)=F(-u)$, and the $x$-dependence at the limits of integration has been omitted due to the assumed periodicity over a distance $L$ (which let us take $x=0$ in both limits without loss of generalization). Thus, Eq.~(\ref{eq:sigmak0}) allows to obtain all the eigenvalues for a given correlation function, $F(x-x')$. Note that this equation is the finite size domain version of Eq.~(\ref{eq:chi}) for a discrete spectrum, so that the correlated noise effect is embedded in the discrete spectrum of the Hilbert-Schmidt operator ${\cal Q}$. 

\begin{figure}[htb]
\centering
\includegraphics[width=0.5\columnwidth]{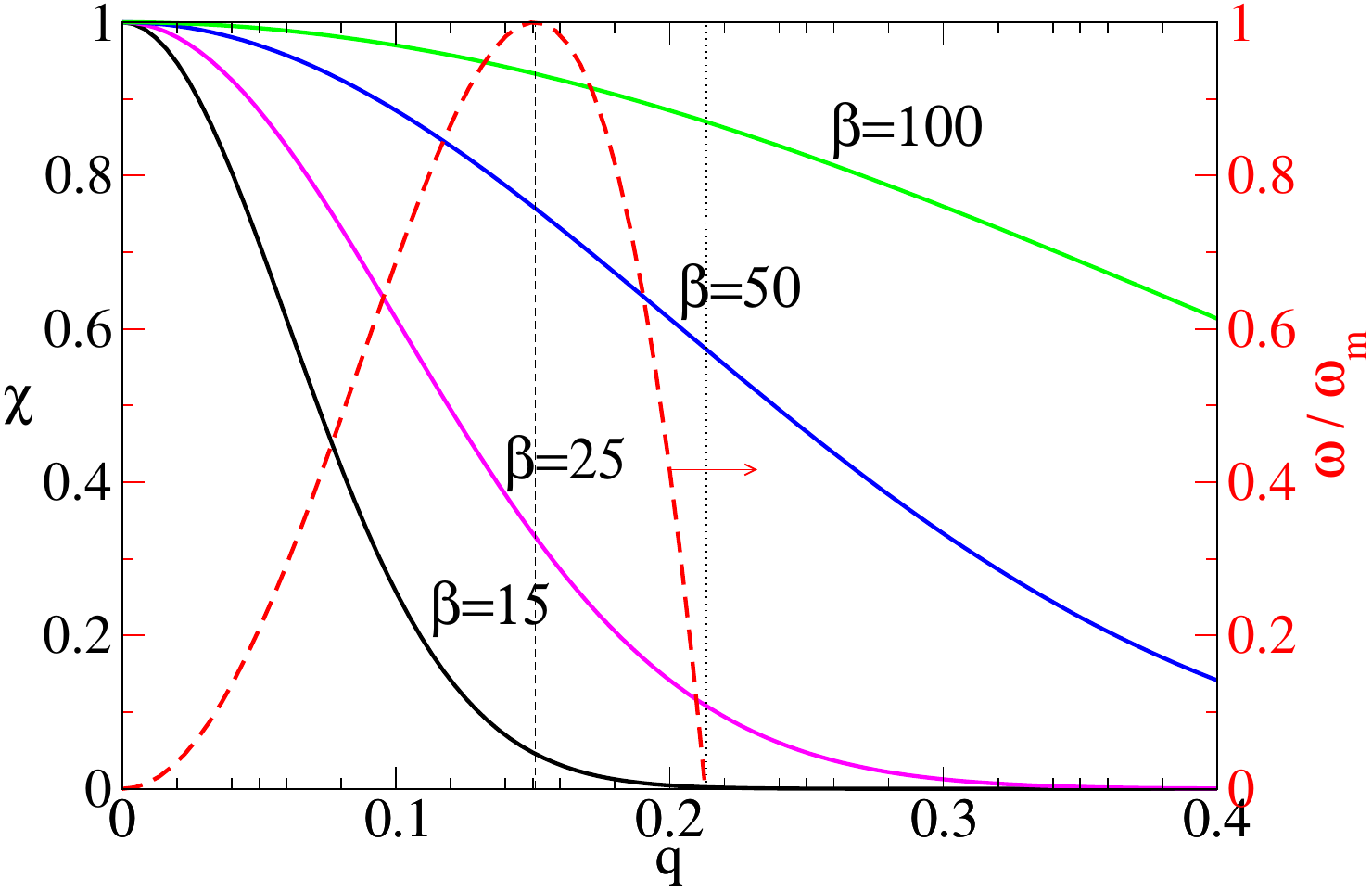}
\caption{
Linear spectrum of eigenvalues for several values of $\beta$ obtained from
Eqs.~(\ref{eq:sigmak2})--(\ref{eq:alpha}). The vertical lines indicate the values of $q_c$ and $q_m$, while the dashed curve corresponds to the deterministic dispersion relation, $\omega(q)$, given by Eq.~(\ref{eq:w_det}).} \label{fig:evalues_q}
\end{figure}

Now, we choose the particular correlation function~\cite{grun_jsp06}
\begin{equation}
 F(u,\ell_c)=\left\{
 \begin{tabular}{lr}
  $Z^{-1} \exp \left[
  -\frac {1}{2} \left( \frac {L}{\ell_c} \sin \left( \frac {\pi u}{L} \right) \right)^2 \right], $ & $\ell_c>0$ \\
  $\delta(u)$, & $\ell_c=0$ ,
 \end{tabular}
 \right.
 \label{eq:F}
\end{equation}
where $\ell_c$ is the correlation length, and $Z$ is such that $\int_0^L
F(u,\ell_c) du=1$. As shown in Appendix~\ref{app:eigenv} we find that the
eigenvalue in Eq.(\ref{eq:sigmak0}) becomes
\begin{equation}
 \widehat F(q_k)=\chi_k=\frac{I_k(\alpha)}{I_0(\alpha)},
 \label{eq:sigmak2}
\end{equation}
where
\begin{equation}
 \alpha= \left( \frac {L}{2 \ell_c} \right)^2=\beta^2.
 \label{eq:alpha}
\end{equation}  
We show in Fig.~\ref{fig:evalues_q} this eigenvalues spectrum for several values of $\beta=L/(2 \ell_c)$. Note that for $\beta \rightarrow \infty$ (i.e. $\ell_c \rightarrow 0$), we have $\chi_k \rightarrow 1$ for all $k$, which leads to the limiting case of white (uncorrelated) noise. For decreasing $\beta$ (larger $\ell_c$'s) the width of the spectrum curve diminishes monotonously. The effect of the correlation region (i.e. not negligible values of $\chi_k$) on the film instability can be put in evidence by comparing it with the dispersion relation $\omega(q)$ as given by the deterministic LSA, Eq.~(\ref{eq:w_det}) (see dashed line in Fig.~\ref{fig:evalues_q}). For $\beta<4.16$, all modes (stable and unstable ones) are affected by the noise with increasing effect on stable ones as $\beta$ increases. On the other hand, for $\beta>4.16$ only unstable modes are affected by the thermal noise. Note that this limiting value is related to the value of $\ell_c$, so that the periodicity of the problem, $L$, and the wavelength of maximum growth, $\lambda_m$, plays a role to determine these regions.

\begin{figure}[htb]
\centering
\subfigure[$t=200$]
{\includegraphics[width=0.45\columnwidth]{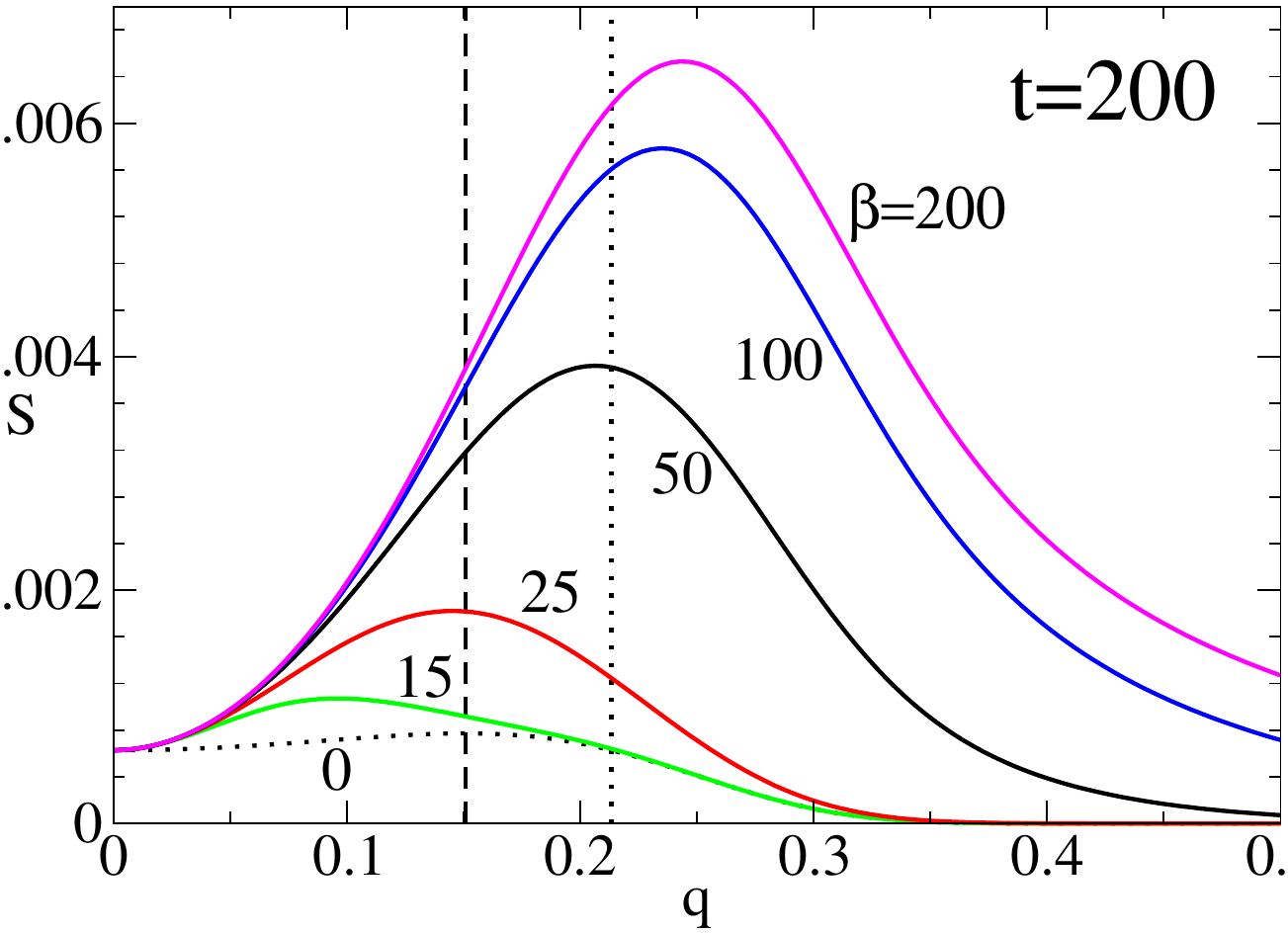}}
\centering
\subfigure[$t=2000$]
{\includegraphics[width=0.45\columnwidth]{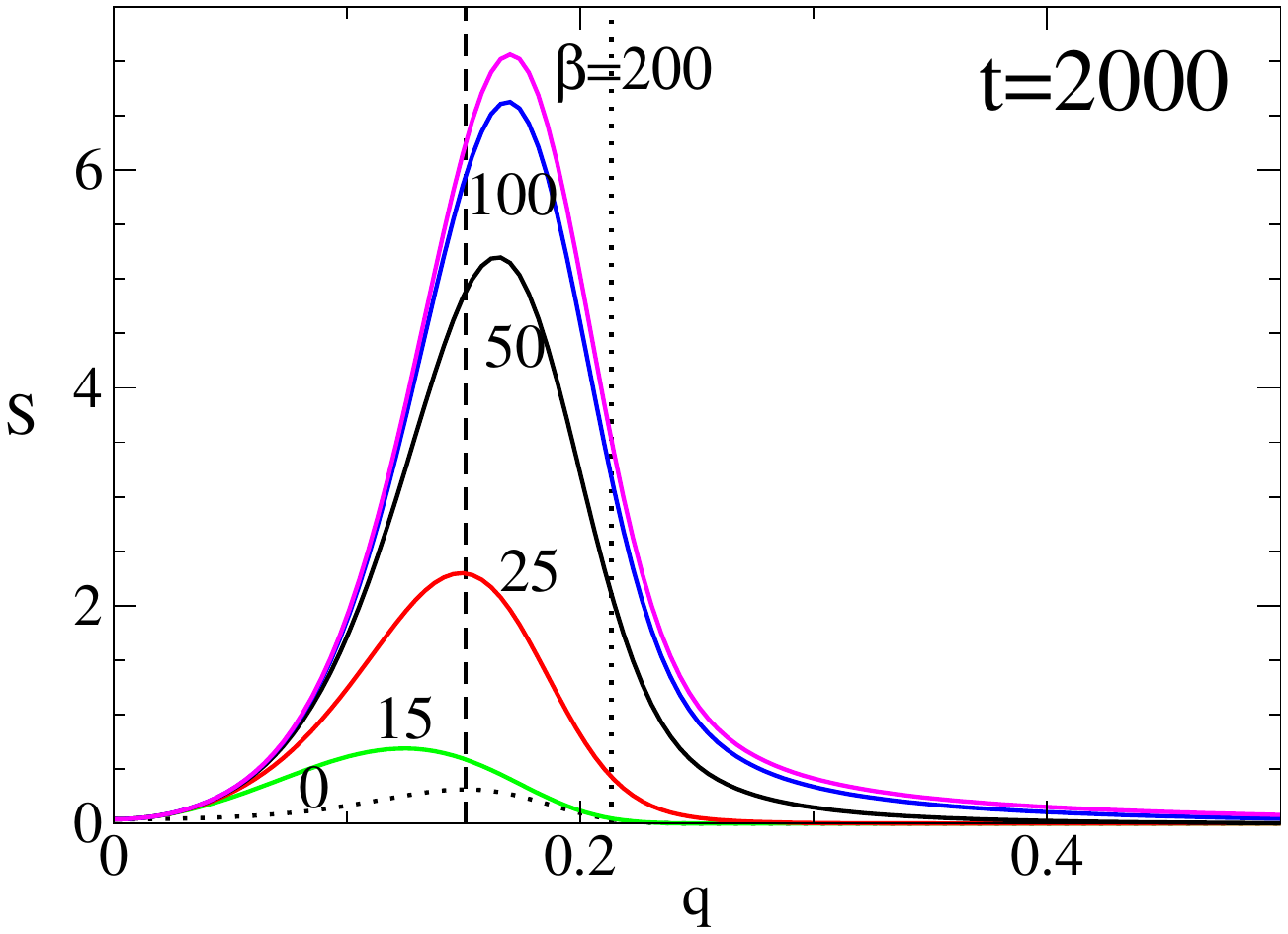}}
\caption{
Power spectrum at two different times for $\sigma =5\times 10^{-5}$, and several values of $\beta$ ($L=500$ and $\ell_c=16.67,10,5,2.5,1.25$) as given by the linear prediction in Eq.~(\ref{eq:S}) and the initial perturbation in Eq.~(\ref{eq:delta0}). The vertical dashed and dotted lines correspond to the wavenumber of maximum growth rate ($q_m=2 \pi/\lambda_m$) and marginal stability ($q_c=2 \pi/\lambda_c$), respectively. The dotted curve shows the deterministic spectra ($\sigma=0$).}
\label{fig:Ssto_times}
\end{figure}

The actual effect of $\ell_c$ on the evolution of the instability is clearly observed in the power spectrum of the perturbation, $S(q,t)$, as predicted by the linear stability analysis in Section~\ref{sec:LSA}. Figure~\ref{fig:Ssto_times} shows $S$ versus $q$ at $t=200$ and $t=2000$ as given by Eqs.~(\ref{eq:S}) ($t=t'$) and (\ref{eq:delta0}). As expected from the analysis of Fig.~\ref{fig:evalues_q}, the inclusion of stochastic noise increases the amplitude of the modes $q>q_c$ (dotted vertical line) which are otherwise stable in the deterministic case. Note that $\beta=0$ (i.e. $\ell_c=\infty$) is coincident with this case in which $\sigma = 0$. This increment increases with $\beta$, that is as the type of noise becomes closer to white noise ($\ell_c \rightarrow 0$). Taking very large values of $\ell_c$, i.e. of $\beta$ (e.g. $\beta=50$), is equivalent to disregarding the noise at all ($\sigma=0$), at least for short wavelengths, since both spectra are practically coincident for early times, and only differ at later times for smaller $q$'s.

\begin{figure}[htb]
\centering
\subfigure[$\beta=15$]
{\includegraphics[width=0.45\columnwidth]{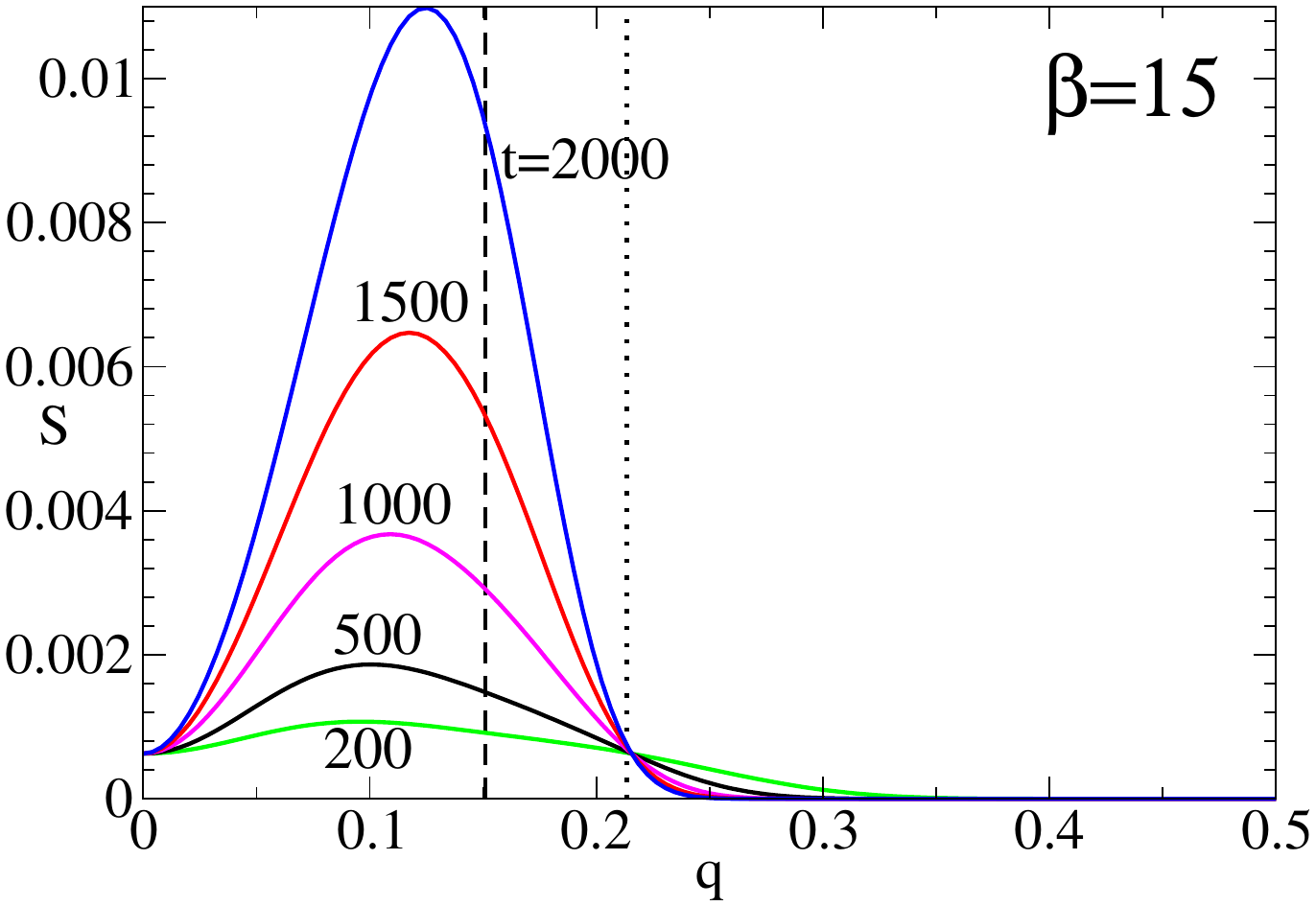}}
\centering
\subfigure[$\beta=50$]
{\includegraphics[width=0.45\columnwidth]{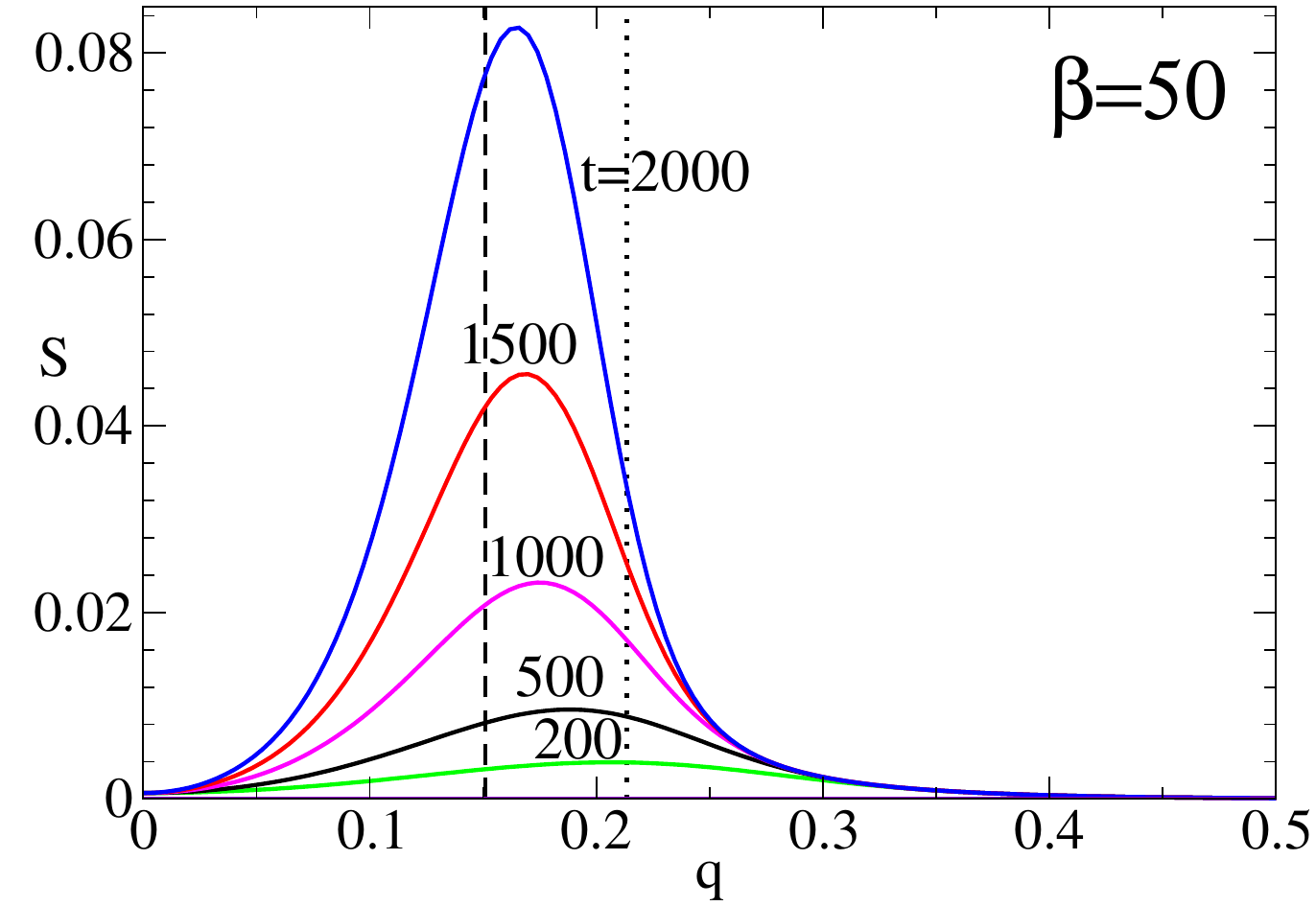}}
\caption{
Power spectrum for two different values of $\beta$ for $\sigma =5\times
10^{-5}$, and several times. (a) $\beta= 15$ ($\ell_c= 16.67$), and (b) $\beta=
50$ ($\ell_c= 50$).}
\label{fig:Ssto_betas}
\end{figure}

\begin{figure}[htb]
\centering
\includegraphics[width=0.45\columnwidth]{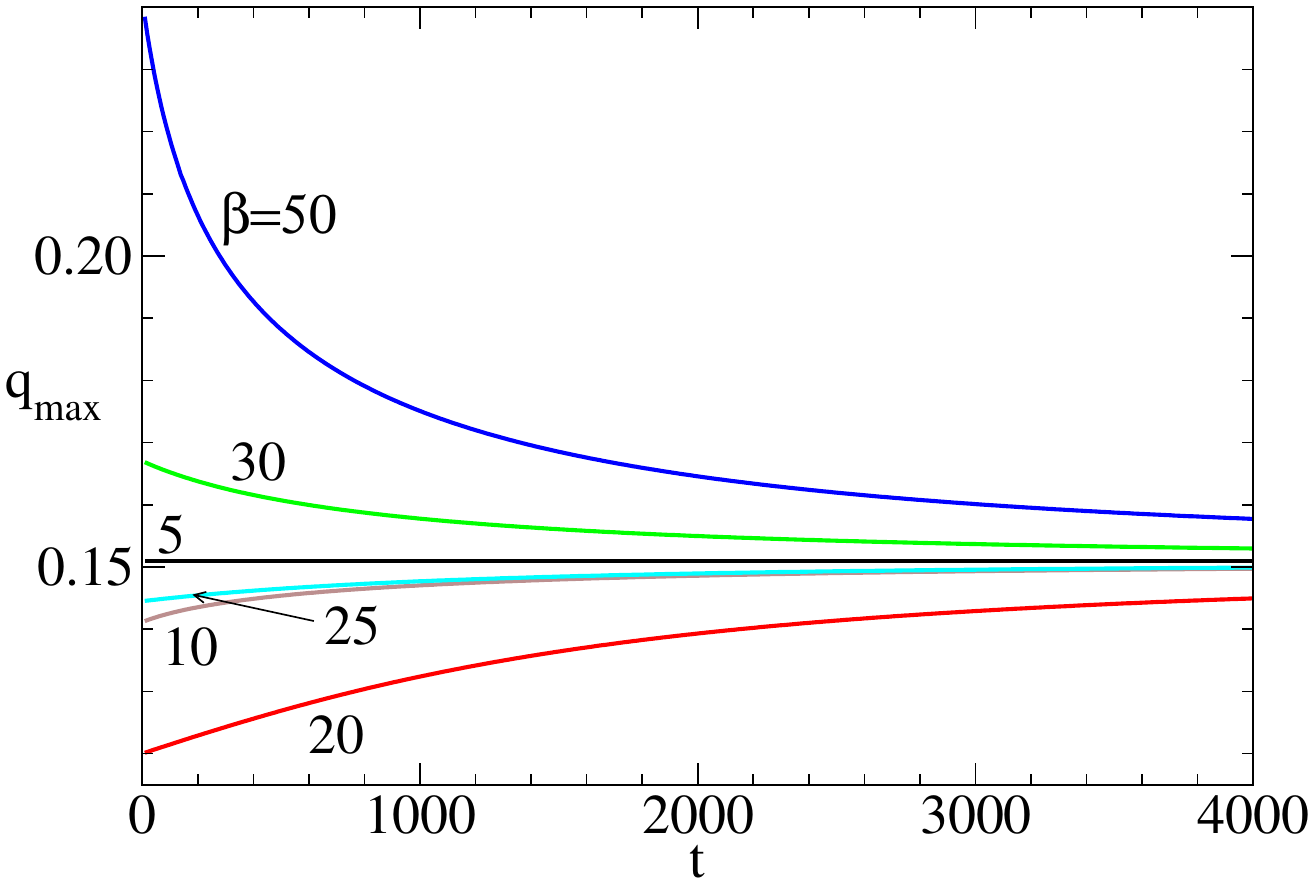}
\caption{
Time evolution of the wavenumber of the maximum of the spectra, $k_{max}$, for different values of $\beta$.}
\label{fig:kmax_t}
\end{figure}

In Fig.~\ref{fig:kmax_t} we show the time evolution of the wavenumber of the maximum of the spectra, $q_{max}(t)$, for different values of $\beta$. Note that for small $\beta$ (say $\beta <5$), we find $q_{max} \approx q_m$ in agreement with the deterministic prediction. As $\beta$ increases up to $\beta \approx 20$ we find that $q_{max}<q_m$ and that it approaches $q_m$ from below. For $\beta \gtrsim 20$ the initial behaviour of $q_{max}$ becomes closer to $q_m$. Finally, for $\beta \gtrsim 27$, $q_{max} > q_m$ for all time, and it approaches $q_m$ from above.

\section{Numerical implementation in a finite domain}
\label{sec:num}

In order to understand the nonlinear effects in the film instability, we perform numerical simulations of the evolution of the film governed by the nonlinear Eq.~(\ref{eq:h1_ad}). The calculations are carried out in a computational domain defined by $0 \leq x \leq L$, which is divided into cells of size $\Delta x$ (typically, we use $\Delta x= 0.1=h_{\ast}$ which assures convergence of the numerical scheme~\cite{DKB01}).

Equation~(\ref{eq:h1_ad}) is discretized in space using a central finite difference scheme. Regarding the spatial dependence of the noise term, we use here only the sinusoidal modes in Eq.~(\ref{eq:gk}), since no flow boundary conditions are imposed at $x=0,~L$. Time discretization is performed using implicit Crank-Nicolson scheme with relaxation factor equal to $1/2$. Thus, the time evolution of the stochastic term is performed according to Stratonovich rules. We note that all the results presented in this paper are fully converged, as verified by grid refinement; more details about numerical issues can be found in~\cite{DK_jcp02}.  Due to the discretization of the equations, the minimum possible value of the correlation length is $\ell_c=\Delta x$ ($=0.1$ in our case), since the discretized equations cannot make any correlation below this limit. 

To discretize the time-Wiener-processes in the framework of Ito-calculus, we replace $\dot \beta_k(t_n)$ at a time step $t_n$ by the forward difference quotient
\begin{equation}
\dot \beta_k(t_n) \approx \frac {\Delta \beta}{\Delta t_n}= \frac { \beta_k(t_{n+1})-\beta_k(t_{n})}{t_{n+1}-t_{n}}.
\label{eq:dbdt}
\end{equation}
The difference $\Delta \beta$ is normal distributed and the variance is given by the time increment $\Delta t_n$. Thus, we approximate Eq.~(\ref{eq:dbdt}) by
\begin{equation}
\frac {\Delta \beta}{\Delta t_n}= \frac { {\cal N}_k^n}{\sqrt{\Delta t_n}},
\label{eq:dbdtap}
\end{equation}
where ${\cal N}_k^n$ is a computed generated random number which is approximately $N(0,1)$-distributed, i.e. its histogram is close to a Gaussian with media zero and unity standard deviation (we used the GASDEV routine from Numerical Recipes~\cite{N-rec}). Altogether, the space-time discrete noise term, Eq.~(\ref{eq:zeta}), is given by
\begin{equation}
 \Theta(x,t)= \frac{1}{\sqrt{\Delta t_n}} \sum_{k=-\frac{N-1}{2}}^{\frac{N-1}{2}} \chi_k {\cal N}_k^n g_k(x),
 \label{eq:xi_xt}
\end{equation}
where $\chi_k$ is given by Eq.~(\ref{eq:sigmak2}), and $g_k(x)$ by Eq.~(\ref{eq:gk}). Thus, Eq.~(\ref{eq:xi_xt}) is used to calculate the noise term in Eq.~(\ref{eq:h1_ad}). 

\begin{figure}[htb]
\centering
\subfigure[ $\sigma =0$, $t=3000$]
 {\includegraphics[width=0.24\columnwidth]{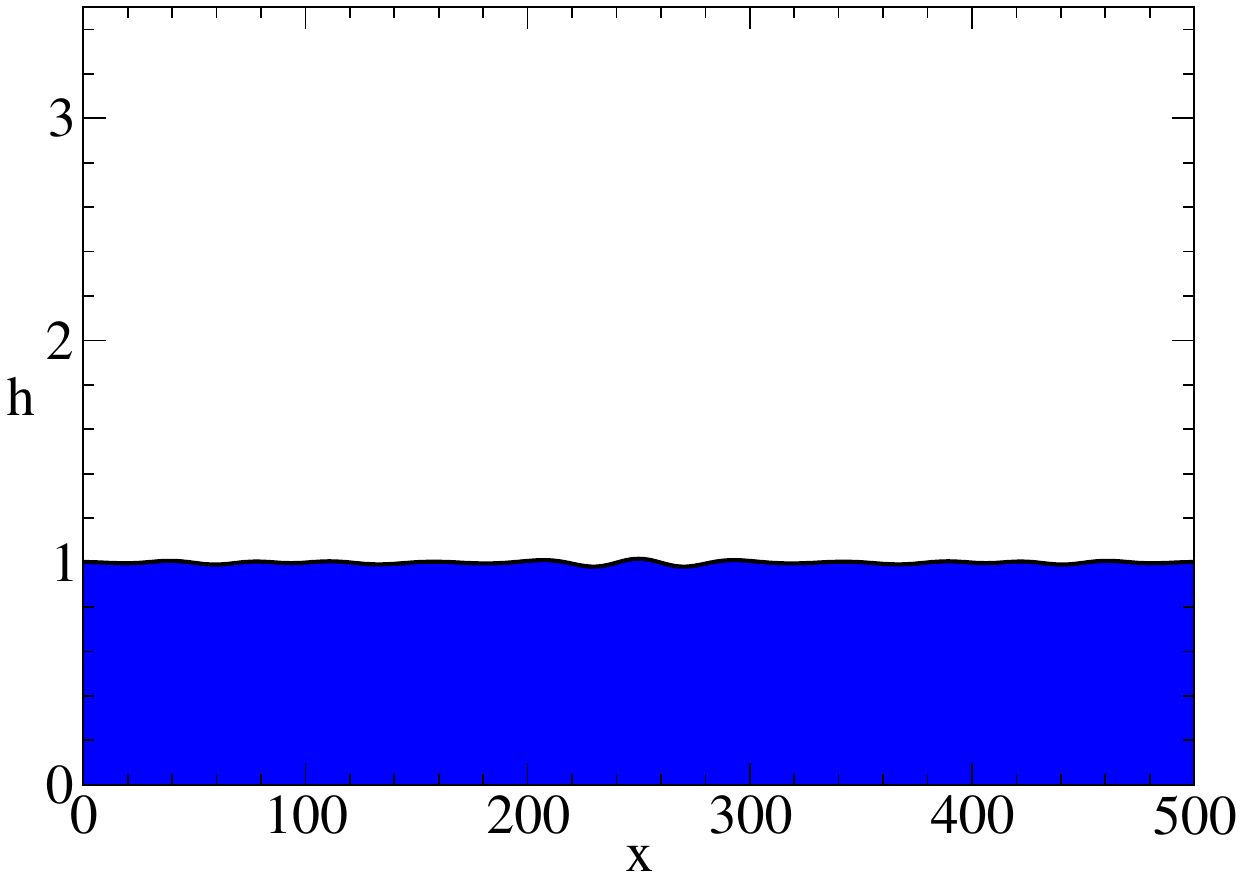}}
\subfigure[ $\sigma =0$, $t=6000$]
 {\includegraphics[width=0.24\columnwidth]{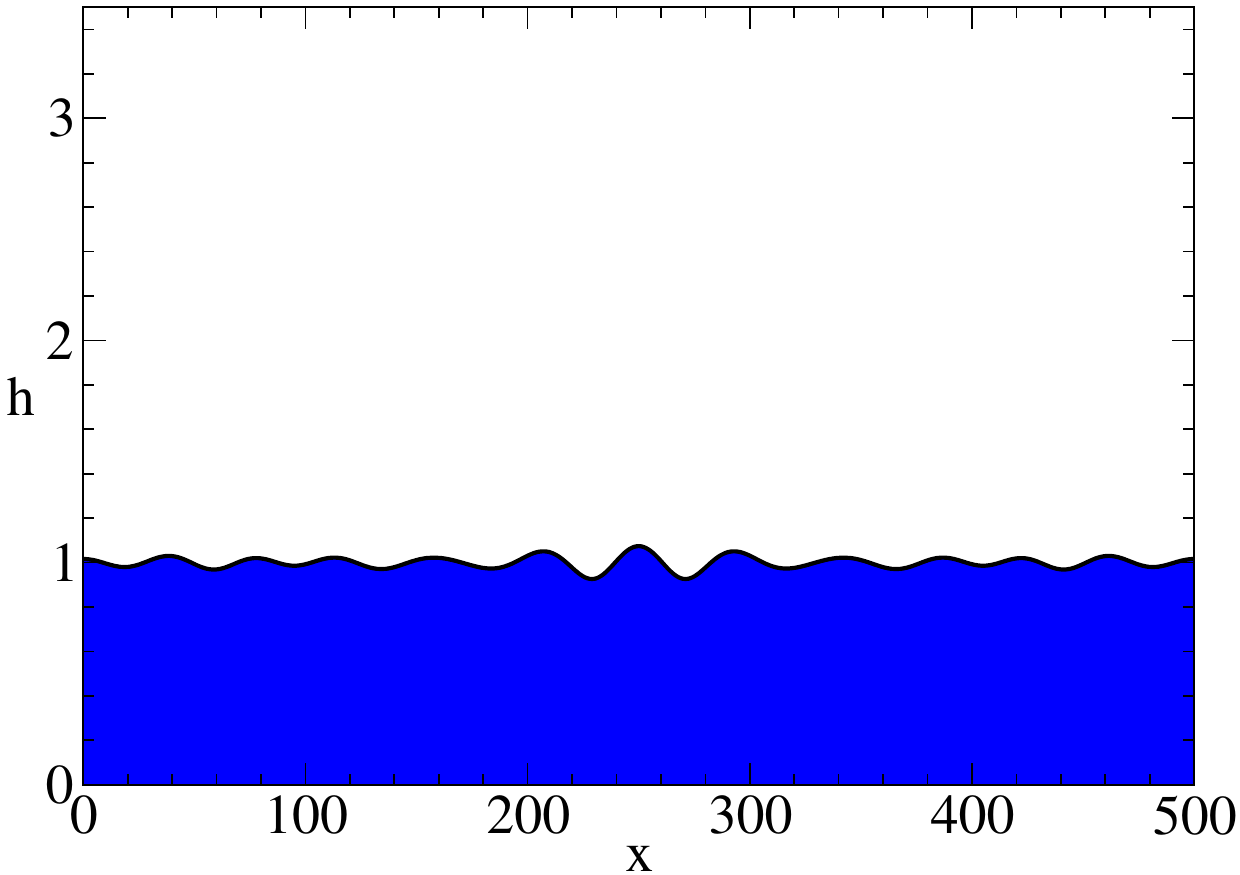}}
\subfigure[ $\sigma =0$, $t=9000$]
 {\includegraphics[width=0.24\columnwidth]{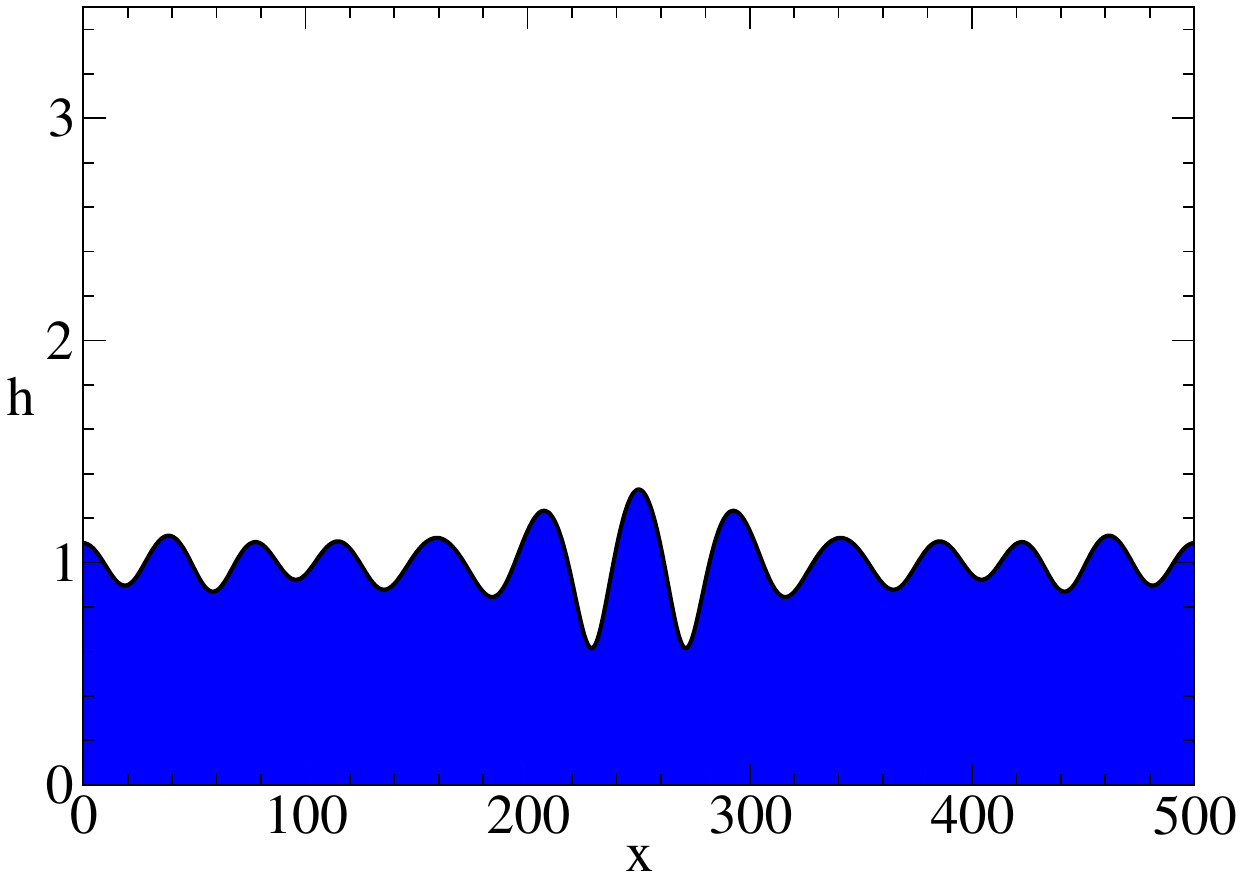}}
\subfigure[ $\sigma =0$, $t=13000$]
 {\includegraphics[width=0.24\columnwidth]{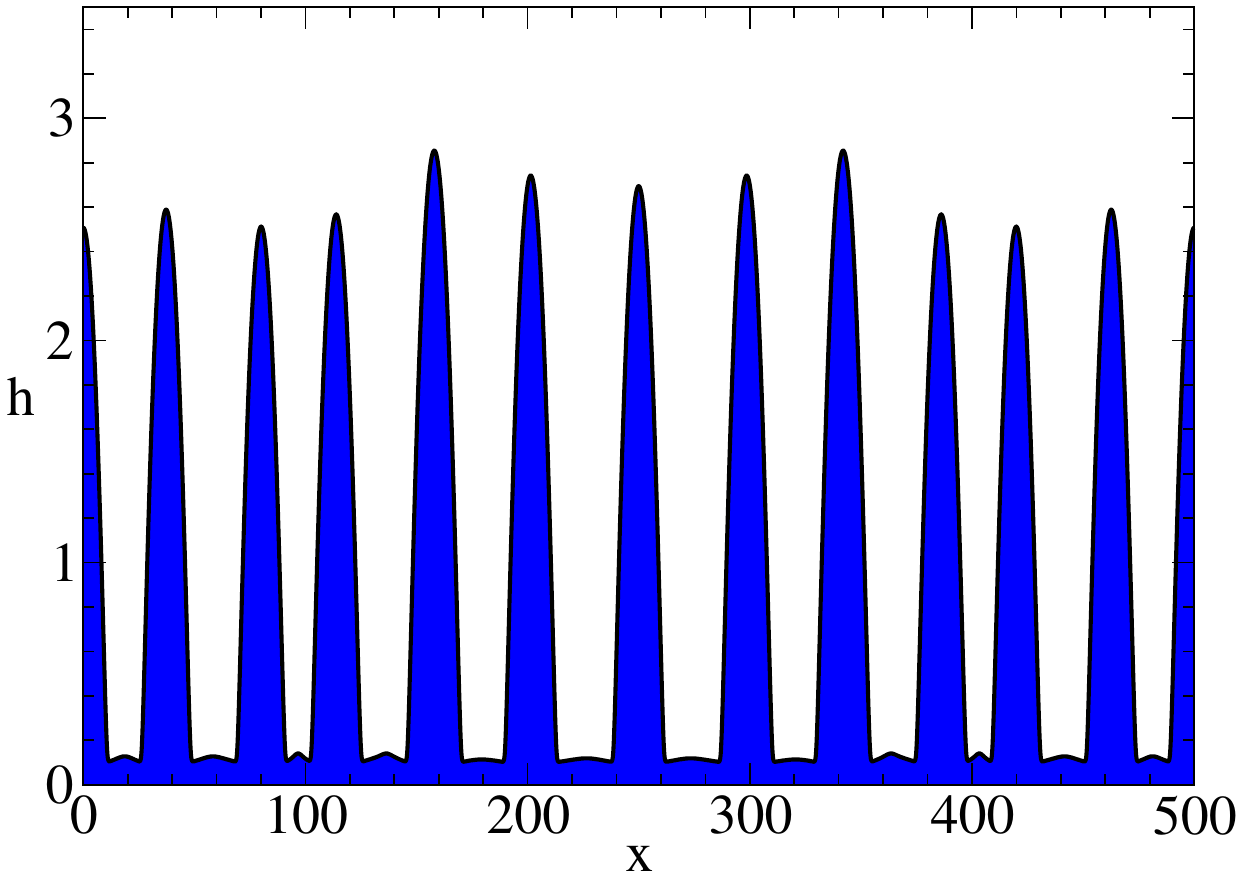}}
\subfigure[$\sigma >0$, $t=3000$]
 {\includegraphics[width=0.24\columnwidth]{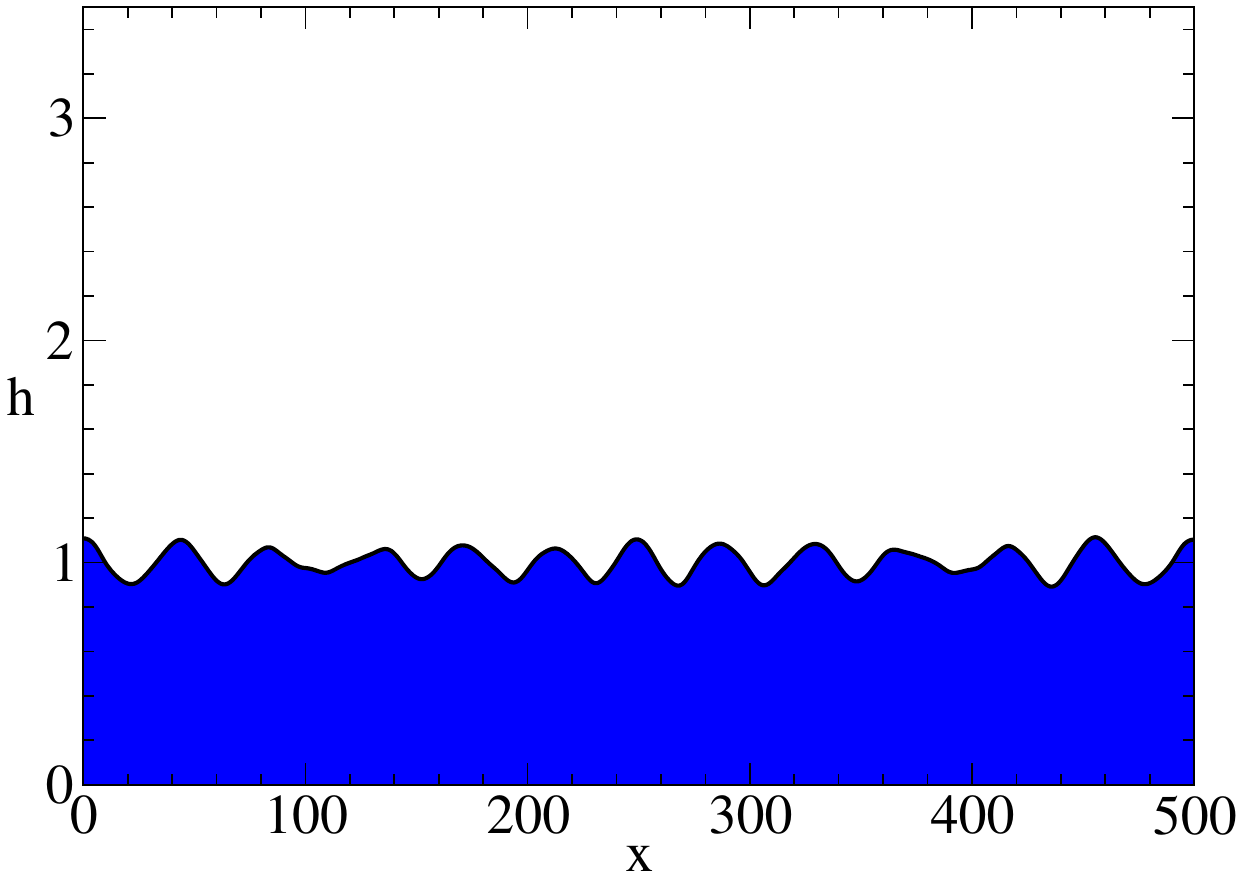}}
\subfigure[$\sigma >0$, $t=6000$]
 {\includegraphics[width=0.24\columnwidth]{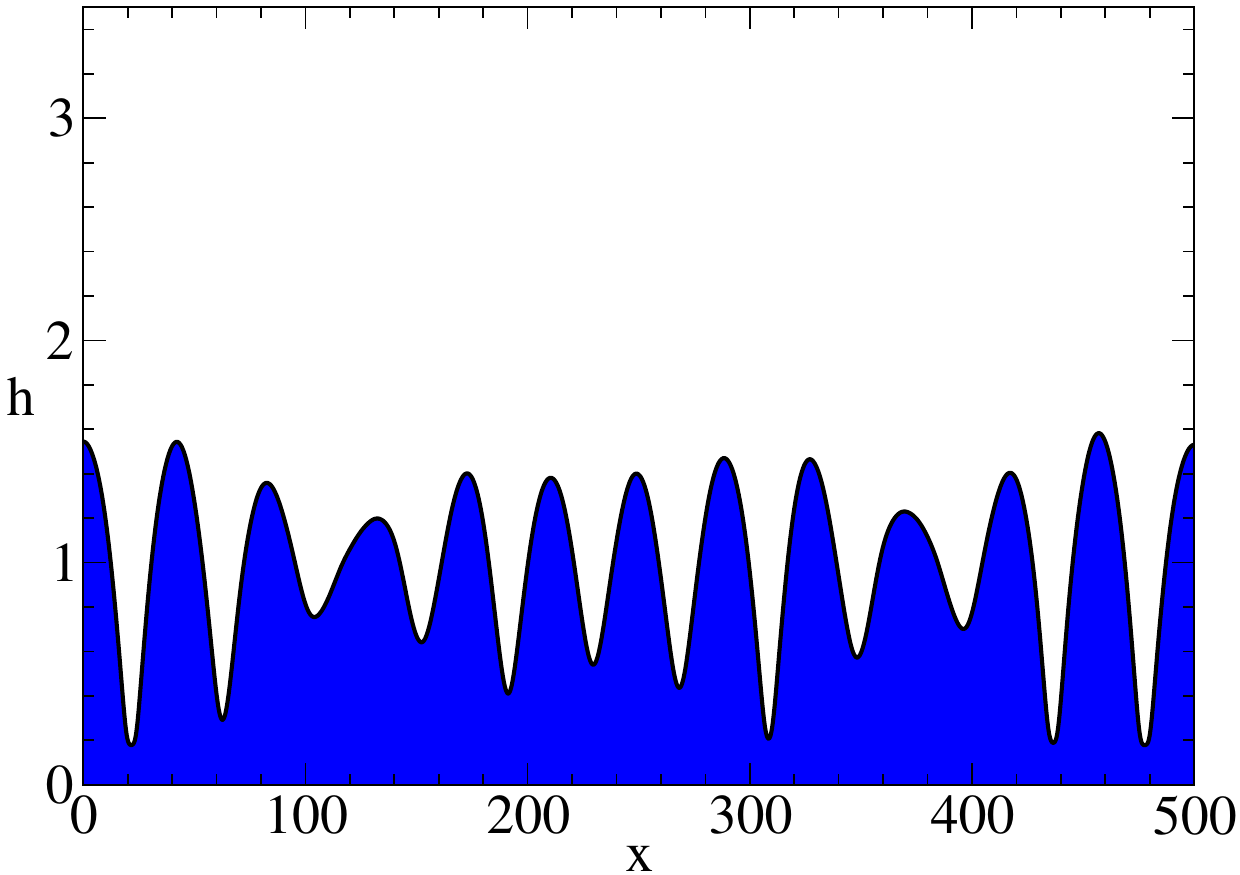}}
\subfigure[$\sigma >0$, $t=9000$]
 {\includegraphics[width=0.24\columnwidth]{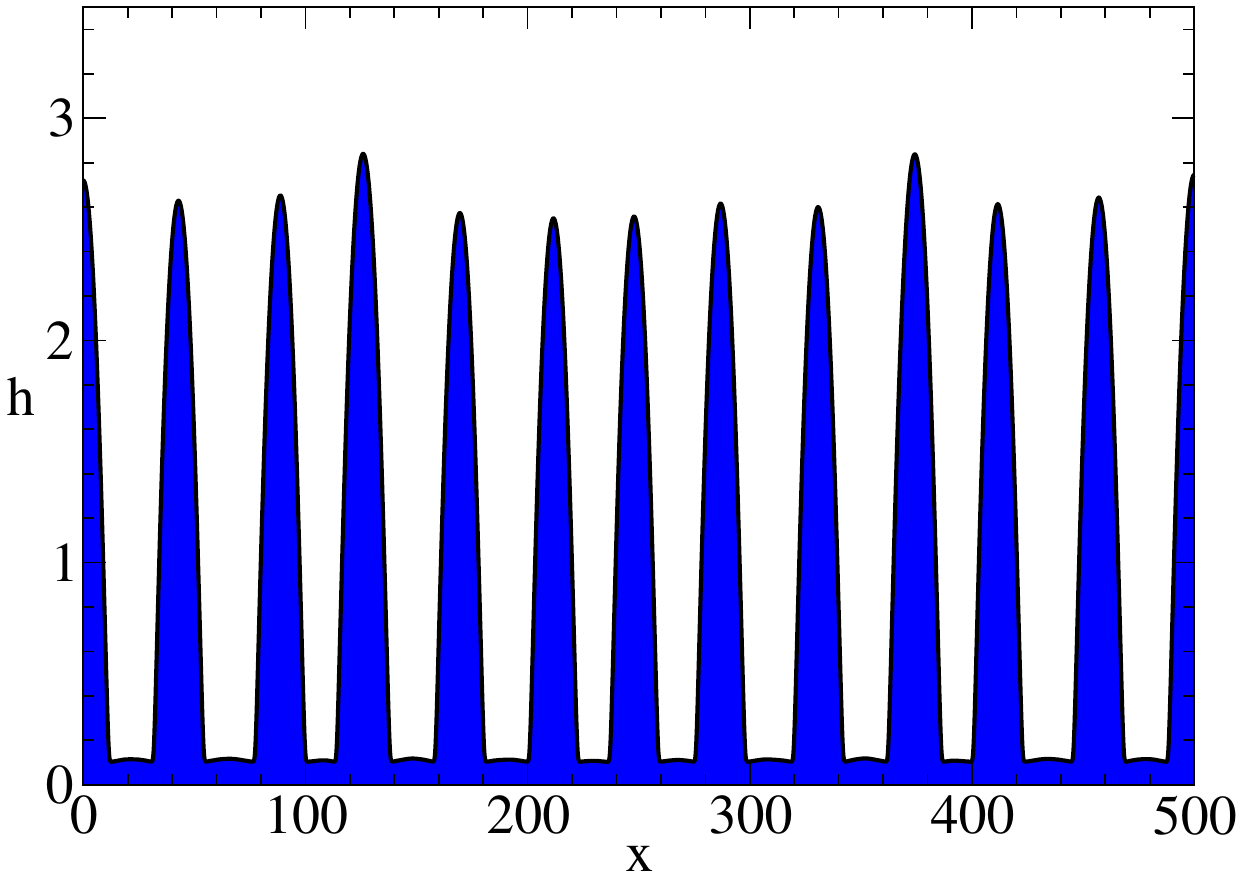}}
\subfigure[$\sigma >0$, $t=13000$]
 {\includegraphics[width=0.24\columnwidth]{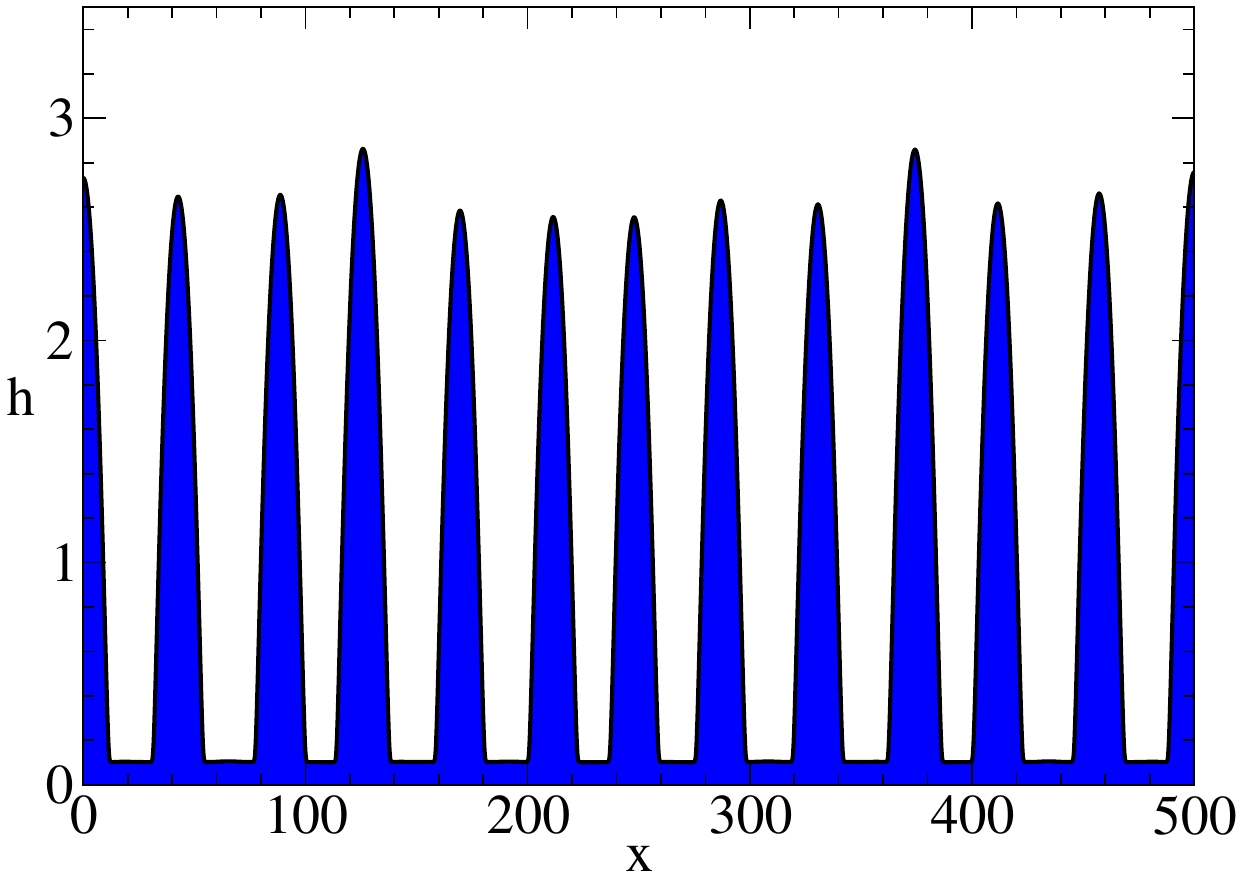}}
\caption{ 
Thickness profiles, $h(x,t)$, of a film at different times: (a)-(d) without noise ($\sigma=0$), and (e)-(h) with noise for a single realization ($\sigma=5\times 10^{-5}$, $\beta=25$, i.e. $\ell_c=10$).}
\label{fig:hx_01}
\end{figure}

Each realization of the stochastic process requires a given seed for ${\cal N}$. Then, some of the numerical results presented below correspond to a single realization, and others to the average of $20$ realizations (different seeds). A typical example of the evolution of a film with and without noise effects for a single realization (i.e. a given seed) is shown in Fig.~\ref{fig:hx_01}. Note that for the same time of the evolution, the amplitudes of the corrugations are much larger for $\sigma>0$ (Fig.~\ref{fig:hx_01}e-h) than for $\sigma=0$ (Fig.~\ref{fig:hx_01}a-d). Thus, one of the effects of the noise is to decrease the duration of the breakup process. 

\begin{figure}[ht]
\centering
\subfigure[]
 {\includegraphics[width=0.48\columnwidth]{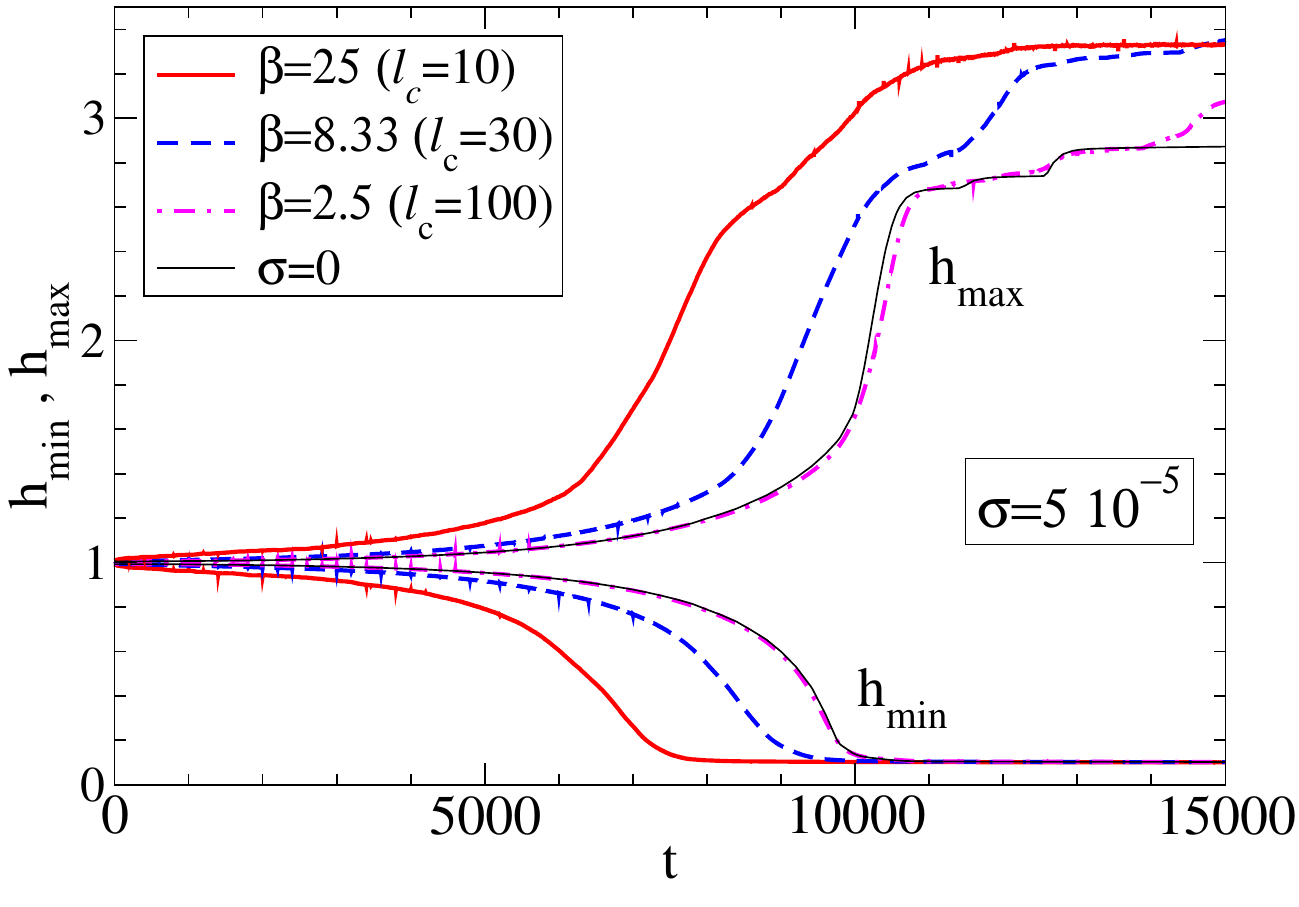}}
\subfigure[]
 {\includegraphics[width=0.48\columnwidth]{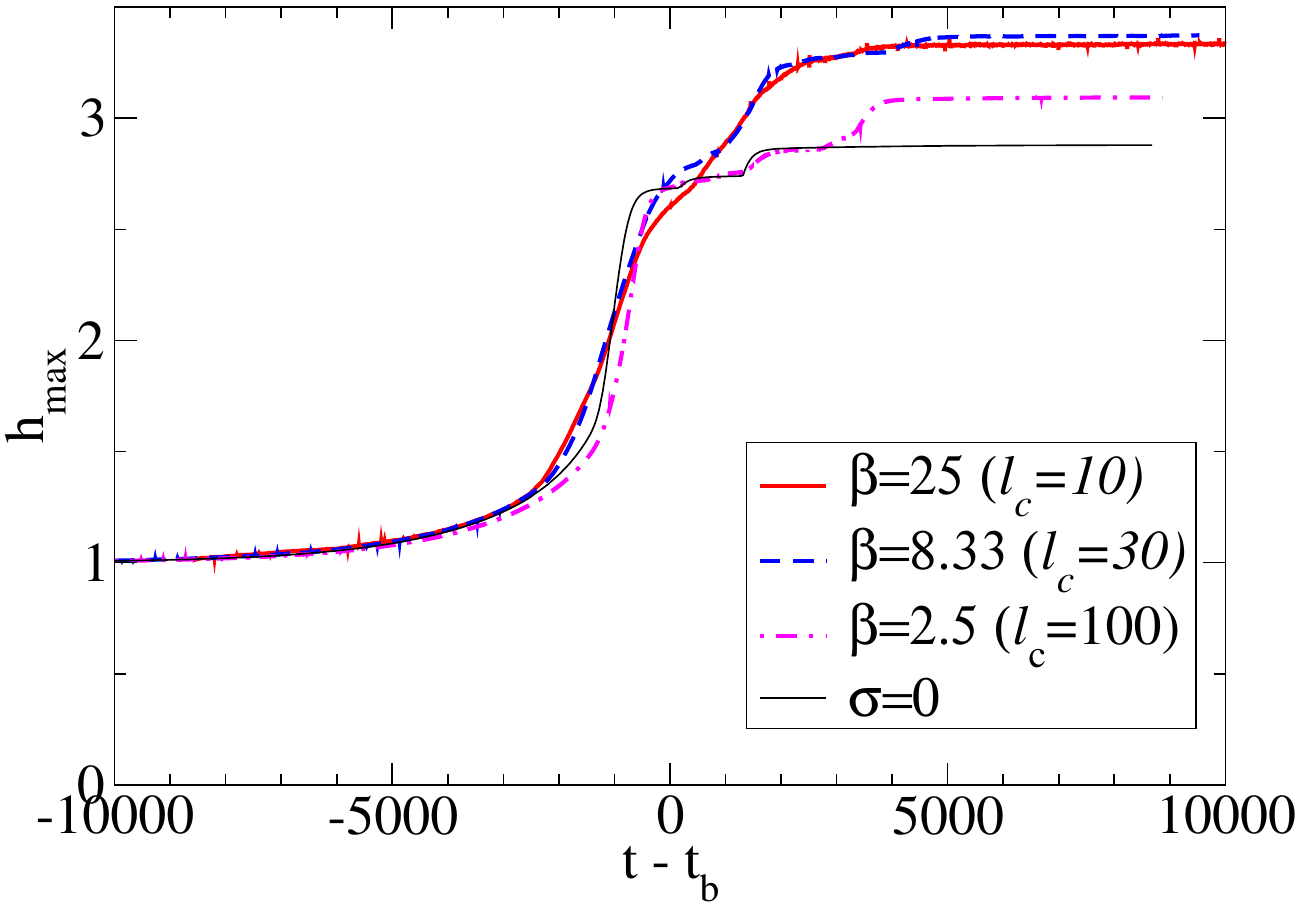}}
\caption{
(a) Maximum, $h_{max}$, and minumum, $h_{min}$, thicknesses versus time for several values of $\ell_c$ averaged over $20$ realizations. (b) Maximum thickness, $h_{max}$, versus the shifted time $t-t_b$, where  $t_b= 8600$, $10468$, and $11122$ for $\beta= 25$, $8.33$, and $2.5$ ($\ell_c=10$, $30$, and $100$.)}
\label{fig:hmax}
\end{figure}

In order to study how the correlated noise affects the time evolution of the instability we first concentrate on the time it takes to appear the first rupture of the film. By first rupture time, we mean the moment when the film first reaches its possible smallest value, which is $h_\ast$. Figure~\ref{fig:hmax}a shows the time evolution of the minimum of $h(x,t)$, namely $h_{min}(t)$. Clearly, as $\ell_c$ increases the breakup time, $t_b$, increases, such that $\beta=2.5$ ($\ell_c=100$) is practically coincident with the case without noise ($\sigma=0$), which has the largest time. For $\sigma>0$, this time decreases for increasing $\sigma$. 

A parameter of interest for the drop formation problem after the first breakup is the evolution of the maximum thickness as the final static configuration is reached. In Fig.~\ref{fig:hmax}a we show the average of $h_{max}(t)$ over $20$ realizations for different values of $\beta$. We also plot $h_{min}(t)$ for reference, and define the corresponding breakup times, $t_b$, as $h(t_b)=1.05\, h_{\ast}=0.0105$. Figure~\ref{fig:hmax}b shows that in fact the evolution of $h_{max}(t)$ is very weakly dependent on $\beta$ (i.e. $\ell_c$), since the curves $h_{max}$ versus $t-t_b$ are practically superimposed. This result implies that the noise does not have any effect on the drop formation process after the breakup of the film, that is, during the dewetting stage following the pinch off.

\begin{figure}[ht]
\centering
\subfigure[$\beta=25$, early times]
{\includegraphics[width=0.49\columnwidth]{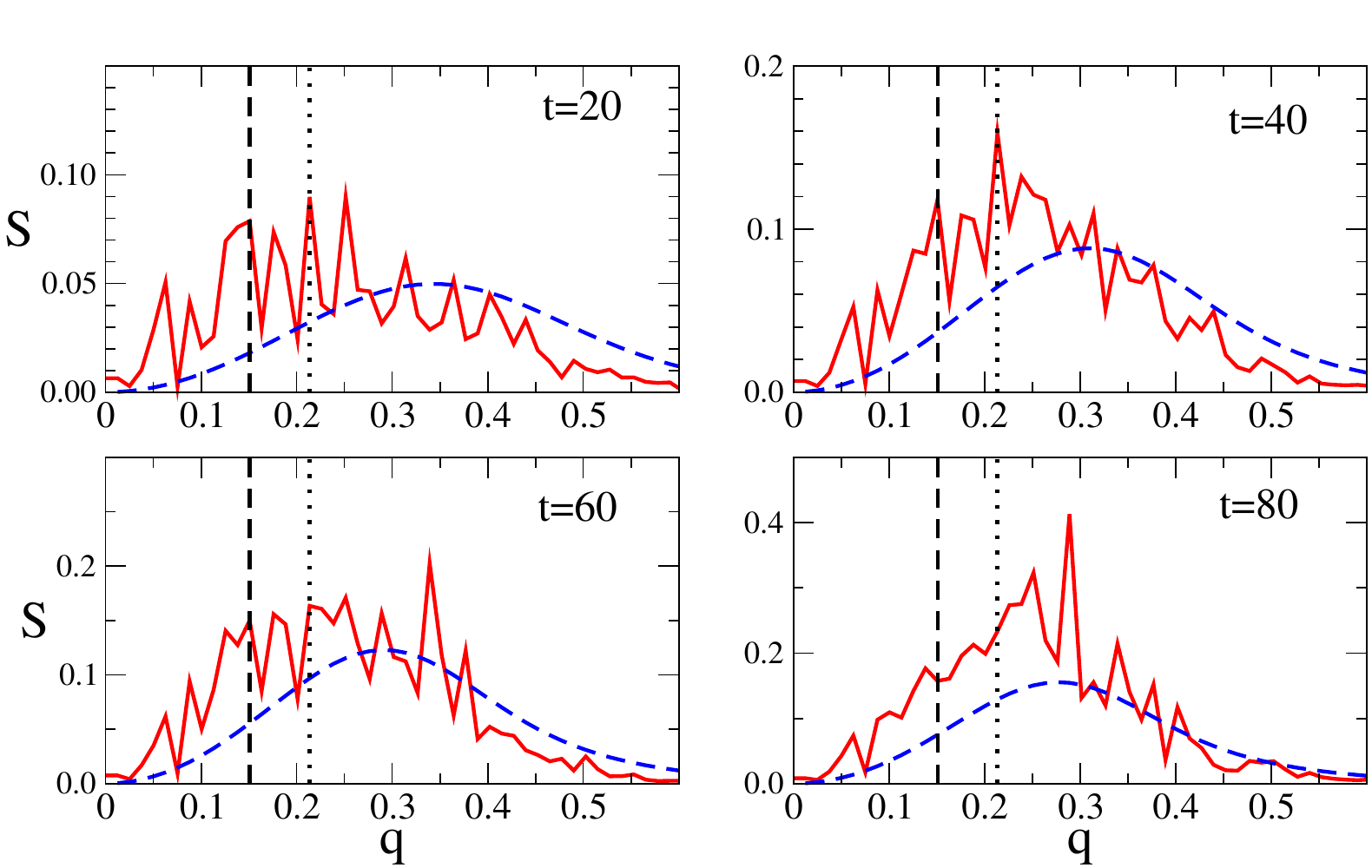}}
\subfigure[$\beta=25$, late times]
{\includegraphics[width=0.49\columnwidth]{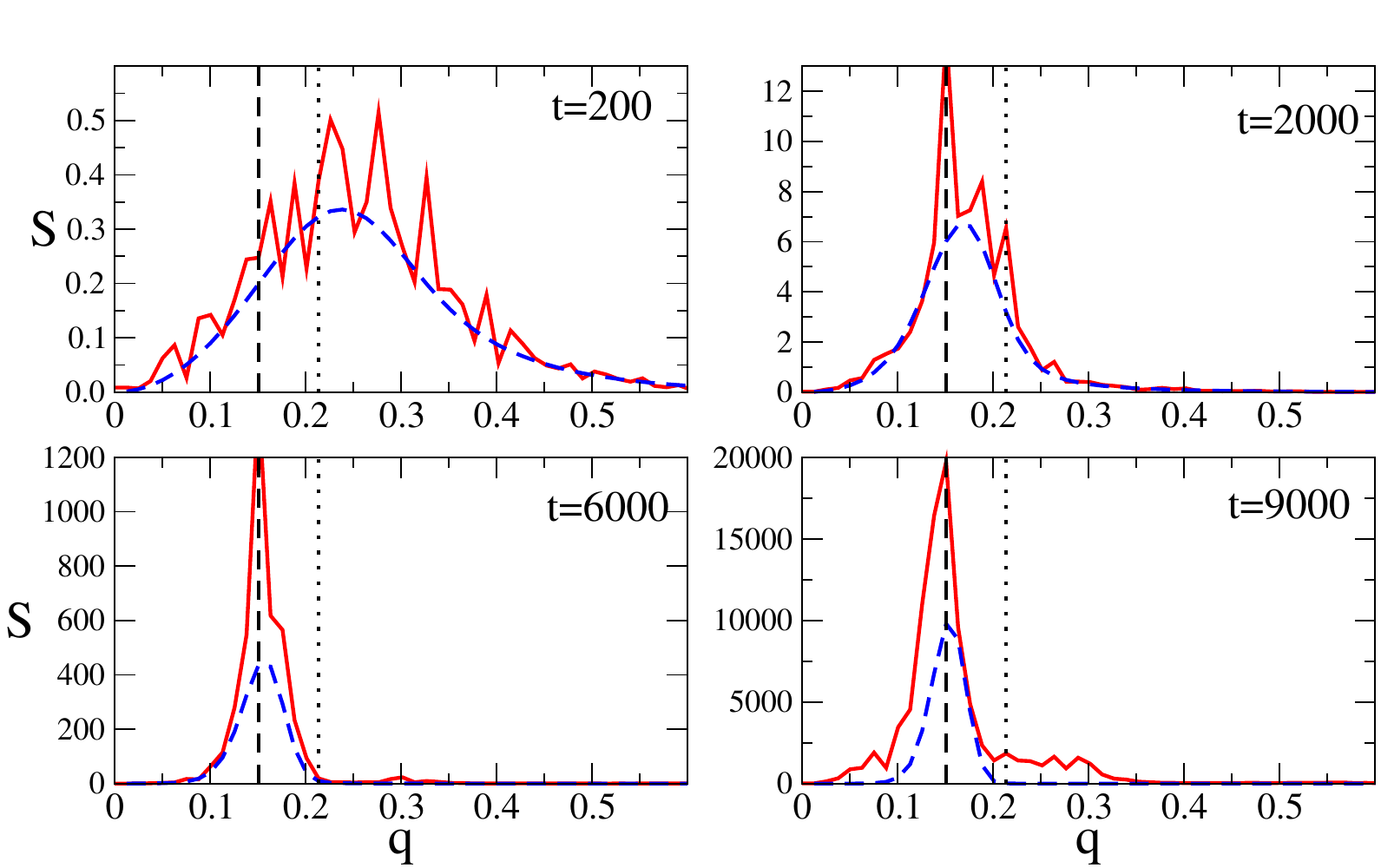}}
\caption{
Numerical power spectra, $S(q,t)$ (solid lines), for (a) \emph {early} and (b) \emph {late} times for $\sigma =5\times 10^{-5}$ and $\ell_c=10$ ($\beta=25$) averaged for $20$ realizations of the problem defined in Fig.~\ref{fig:hx_01}a. The dashed lines are the corresponding predictions of the LSA. The vertical dashed line corresponds to the wavenumber of maximum growth in the deterministic case, $q_m=0.151$, while the dotted one to the marginal value, $q_c=0.215$.}
\label{fig:S_lc10}
\end{figure}

\begin{figure}[ht]
\centering
\subfigure[$\beta=8.33$]
{\includegraphics[width=0.49\columnwidth]{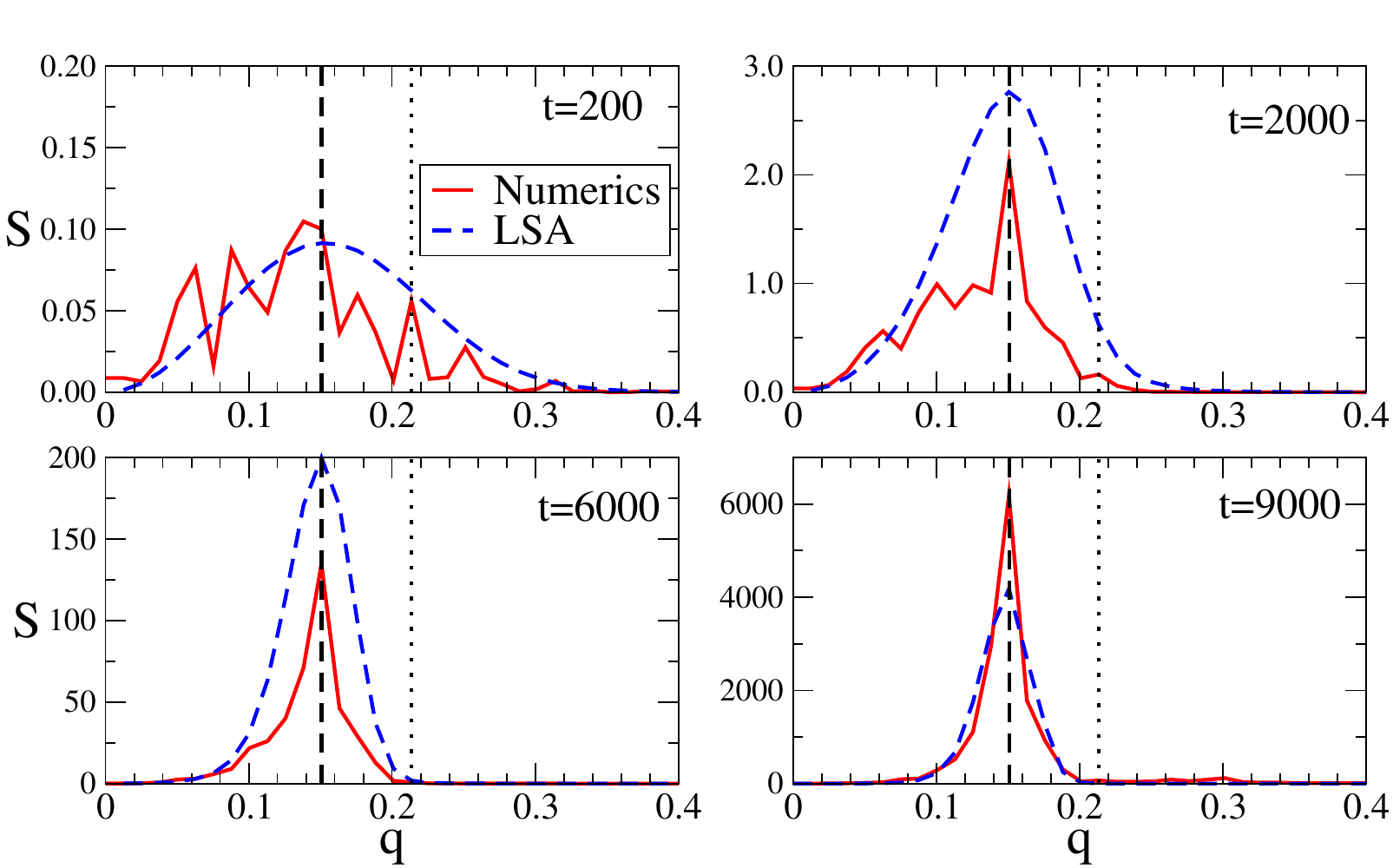}}
\subfigure[$\beta=125$]
{\includegraphics[width=0.49\columnwidth]{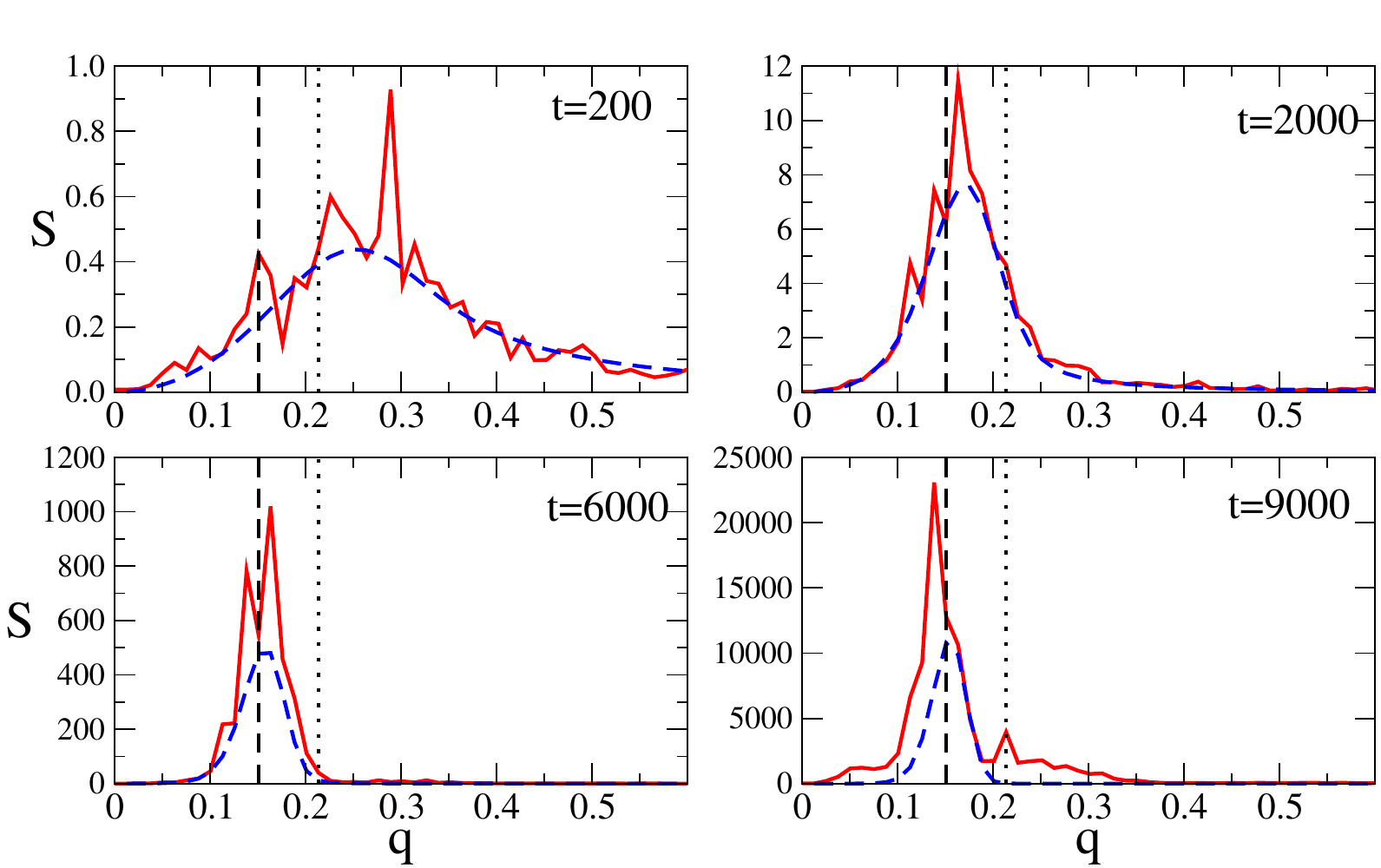}}
\caption{
Numerical power spectra, $S(q,t)$ (solid lines), for $\sigma =5\times 10^{-5}$ and (a)$\ell_c=30$ ($\beta=8.33$) and (b) $\ell_c=2$ ($\beta=125$) averaged for $20$ realizations. The dashed lines are the corresponding predictions of the LSA. The vertical dashed line corresponds to the wavenumber of maximum growth in the deterministic case, $q_m=0.151$, while the dotted one to the marginal value, $q_c=0.215$.}
\label{fig:S_lc30-2}
\end{figure}

Now we aim to study the effects of the correlation length in both linear (early) and nonlinear (late) stages of the instability. To do so, we calculate the Fourier spectra of the thickness profiles for different times. In Fig.~\ref{fig:S_lc10} we show the evolution of the spectra with $\ell_c=10$ ($\beta=25$) for both early and late times. All spectra correspond to an average of $20$ realizations, and no adjusting parameter has been used (the scales for $S$ are different to those used in previous sections because a different normalization was employed in the Fourier transform of the numerical results). For early times, the agreement between numerics and the linear stability prediction, Eq.~(\ref{eq:S}), is very good if one considers that some initial noise is introduced in the numerics. For larger times, the peaks of both spectra approach $q_m$ though the numerics show higher and a bit wider spectra than those predicted by LSA. A similar situation is observed for smaller and larger values of $\beta$ as shown in Fig.~\ref{fig:S_lc30-2}. The main difference is that LSA overestimate the amplitude of the peaks respect to the numerical ones for small $\beta$ (Fig.~\ref{fig:S_lc30-2}a), but the contrary occurs for large $\beta$ (Fig.~\ref{fig:S_lc30-2}b).

\section{Comparison with experiments}
\label{sec:exp}

Previous comparisons between experiments and stochastic models have studied the instability of polymeric films on silicon oxide substrates~\cite{fetzer_prl07,becker_nat03}. However, these comparisons were made without considering spatial correlation, i.e. assuming both spatial and temporal white noise. Also, they mainly employed the integration of the spectra $S(q)$ for all possible values of $q$, and derived quantities from it. Here, instead, we apply the theoretical model described above to experimental results for unstable liquid metal films to evaluate the importance of spatial correlations when considering stochastic instabilities. In order to do this, we do not restrict ourselves to some integrals of the spectra, but employ their complete profiles as a function of the wavenumber, $q$. 

Our experimental data correspond to copper thin films of a few nanometers thick that are melted by the illumination with pulses of an Excimer laser that last some tens of nanoseconds. During these pulses, the metal is in a liquid state, and thus the present hydrodynamic model can be applied. In this configuration, the liquid lifetime of the melted copper is related with the local temperature of the film, i.e. with the spatial distribution of the laser intensity, which spans in a radially symmetric Gaussian profile. After the pulse, the metal solidifies leaving a distinct pattern of holes, drops and/or ridges depending on how long the metal has been in the liquid state. More information about this setup configuration and details on the technique can be found elsewhere~\cite{rack_apl08,kd_pre09,wu_lang10,wu_lang11,fowlkes_nano11}. 

Since the outer regions of the laser spot have shorter liquid lifetimes, one can associate these regions with earlier times of the evolution, and consequently, central regions with later times. Since the laser spot is relatively large, the SEM images of these experiments have the advantage of offering more spatial information than other setups~\cite{fetzer_prl07}. Nevertheless, they have the drawback that the times corresponding to every stage of the evolution are unknown, even if it is possible to order the time sequence in connection with the distance of the image respect to the center of the laser spot~\cite{gonzalez_lang13}. The goal of the following comparison is to show that the experimental observations represented by the spectra require not only a stochastic temporal evolution, but also some spatial correlation in the thermal noise in order to reproduce the full results.

\begin{figure}[hbt]
	\centering
	\subfigure[]
	{\includegraphics[width=0.49\columnwidth]{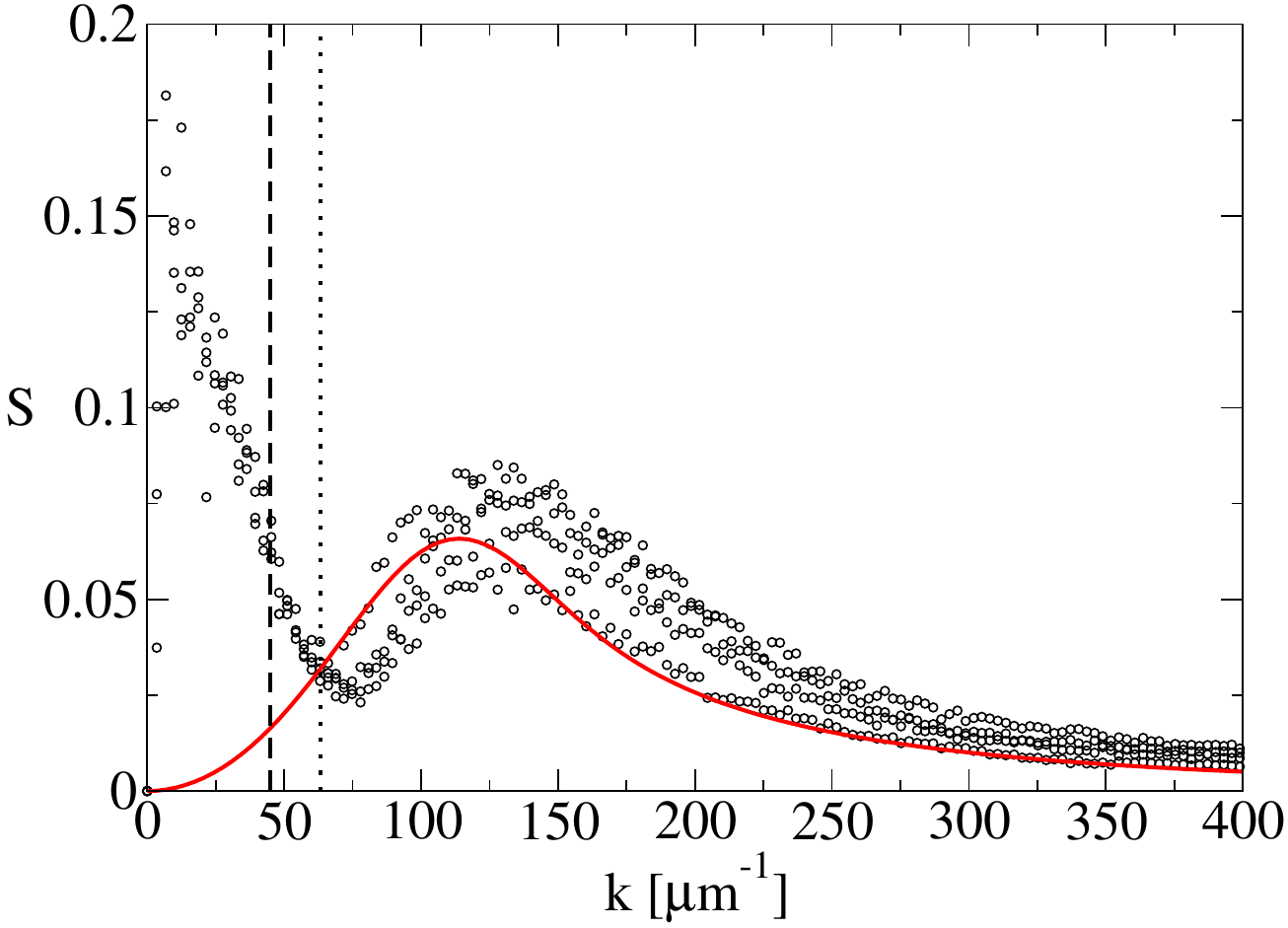}}
	\subfigure[]
	{\includegraphics[width=0.49\columnwidth]{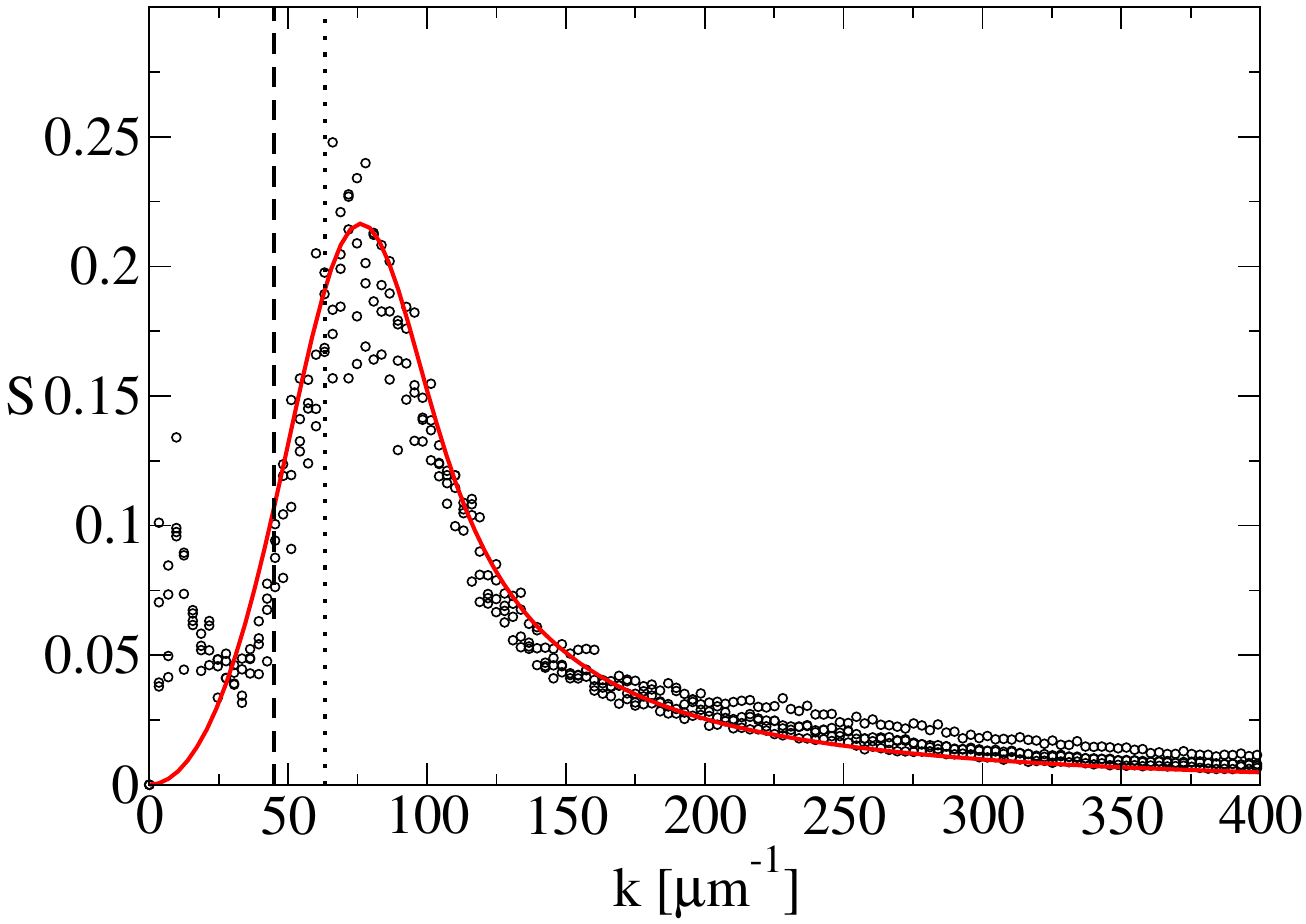}}
	\subfigure[]
	{\includegraphics[width=0.49\columnwidth]{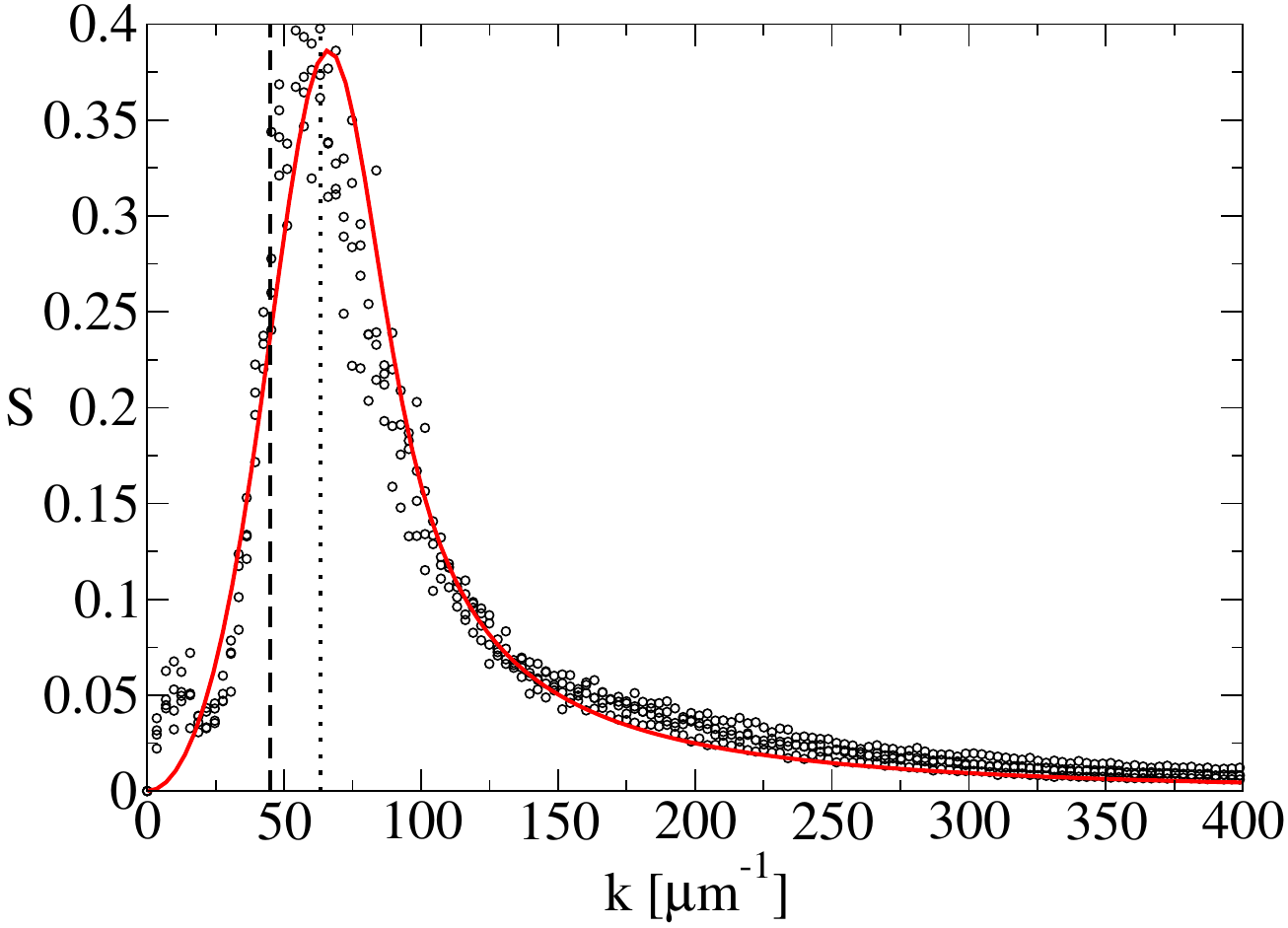}}
	\subfigure[]
	{\includegraphics[width=0.49\columnwidth]{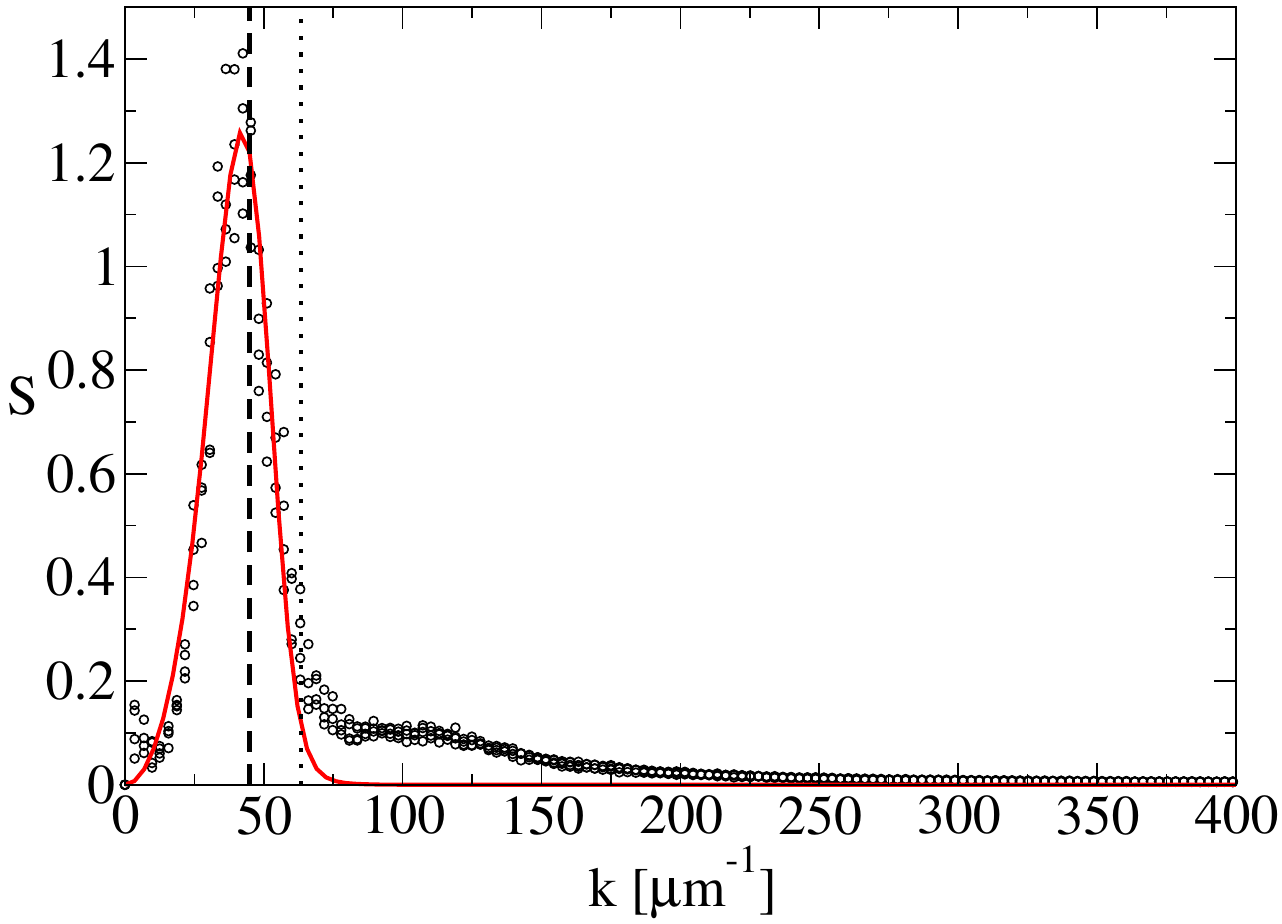}}
	\caption{
		Experimental power spectra, $A_{1D}(k,t)$, (symbols) from Fig. 5 of \cite{gonzalez_lang13}, and theoretical spectra (solid lines) obtained with the present stochastic model with spatial correlation. The experimental spectra are organized in decreasing order of their distance to the center of the laser spot.}
	\label{fig:Sexp}
\end{figure}

In particular, we will concentrate here on the data reported in~\cite{gonzalez_lang13}, where the SEM images of the evolving melted metal were analyzed by using bidimensional (2D) discrete Fourier transform (DFT). Since, the 2D spectra turned out to be radially symmetric in the wavenumber space, $(q_x,q_y)$, the results in Fig. 5 of \cite{gonzalez_lang13} were reported as amplitudes $A_{2D}$, versus $k=(q_{x}^2+q_{y}^2)^{1/2}$. These amplitudes were in fact averaged on circles of radius $k$, and therefore the corresponding 1D amplitude is obtained as $A_{1D}=k A_{2D}^2$ (see symbols in Fig.~\ref{fig:Sexp}). The symbols for both small $k$ and amplitudes ($S < 0.15$) are an artifact of the finite length of the sample in the Fourier calculation.

The parameters for liquid copper are $\gamma =1.304\,~N/m$, and $\mu=4.38\, mPa\,s$. Assuming $T=1500\,~K$ as a typical temperature of the film with thickness $h_0=8~nm$, we have $\sigma= 2.48 \times 10^{-4}$, and $t_0=0.08~ns$. Regarding the intermolecular interaction with SiO$_2$ we use $(n,m)=(3,2)$, $h_{\ast}=0.1\,~nm$  and $A=2.58 \times 10^{-18} J$ (as suggested in~\cite{gonzalez_lang13}). Thus, we have $q_c=63.4\,{\mu m}^{-1}$ and $q_m=44.8\,{\mu m}^{-1}$ (dotted and dashed lines in Fig.~\ref{fig:Sexp}). 

In order to perform the comparison of the experimental and theoretical spectra (see Eq.~(\ref{eq:S})) we choose a constant value for the unknown $\widehat F_0(q)$, namely $\widehat F_0(q)=2 \times 10^{-4}$, and use the same normalization factor for the DFT as in~\cite{gonzalez_lang13}. Thus, we are left only with $t$ and $\beta$ as adjustable parameters. The fitting values for the spectra in Fig.~\ref{fig:Sexp} are given in Table~\ref{tab:exp_vs_theor}. The low local maximum for $k \approx 100 \, {\mu m}^{-1}$ is related to the size of the drops, which is smaller than the distance between them~\cite{gonzalez_lang13}.

Interestingly, we find not only increasing values of time as one moves from inner to outer regions (as expected), but a decrease of the corresponding values of $\beta$ is also required for the fitting. This implies that the stochastic noise is somehow different at the sampled regions which, in turn, correspond to distinct liquid lifetimes. However, the relatively large values of $\beta$ for the first three images suggest that the noise is practically white at the beginning, and that spatial correlation becomes important only for larger times when $\beta$ decreases significantly. In general, it is then expected that the spectrum for earlier times (i.e., near the outer borders of the laser spot) correspond to a quasi white noise, but the noise becomes more and more spatially correlated as one goes to the center of the spot (i.e. as the liquid lifetimes increase). In fact, the correlation length, $\ell_c$, can be estimated considering the value of $\beta$ and the length of the image, which can be assumed as the periodicity length, $L$. For the images corresponding to Fig.~\ref{fig:Sexp} we have $L=2.13~\mu m$, so that we obtain $\ell_c=L/(2\beta)$ as shown in Table~\ref{tab:exp_vs_theor}. Moreover, note that $\ell_c$ finally approaches $\lambda_m$ ($=144\,nm$), which is also very close to $\lambda_m^{exp}$ ($=165\,nm$). Thus, $\ell_c$ turns out to be very close to the average distance between drops. 
\begin{table}[htb]
	\begin{tabular}{ccccccc}
		\hline Fig.~\ref{fig:Sexp} & t(ns) & $\beta$ & $\ell_c$ (nm) & $ \quad \lambda_m/\ell_c  \quad$  & $\quad \lambda_m^{exp}$ (nm) & $\lambda_m^{exp}/\ell_c$ \\ 
		\hline (a) & 0.08 & 175 & 6.1 & 22.9 & 62.8 & 10.3 \\ 
		       (b) & 0.48 & 160 & 6.6 & 21.0& 99.7 & 14.9 \\ 
		       (c) & 0.97 & 140 & 7.6 & 18.4& 125.6 &16.5 \\ 
		       (d) & 6.21 & 7.4 & 144.1 & 0.97 & 165.3 & 1.15 \\  
		\hline 
	\end{tabular} 
	\caption{Best fit values from the comparison of the stochastic model with spatial correlation with experimental spectra of unstable liquid metal films. Here, we have $\lambda_m=144\, nm$.}
	\label{tab:exp_vs_theor}
\end{table}

\section{Summary and conclusions}
\label{sec:conclu}

In this work we have considered the effect of correlated thermal noise on the instability of a liquid thin film under the action of viscous, capillary and intermolecular forces by adding a stochastic term in the lubrication approximation equation for the film thickness. This term depends on the noise amplitude that is spatially self--correlated within a characteristic microscopic distance, $\ell_c$. The linear stability analysis (LSA) of the resulting equation shows that this yields a new factor in the stochastic part of the instability spectrum, which is given by the Fourier transform of the correlation function that can be expressed in terms of the eigenvalues of the Hilbert operator associated with it.

In order to observe the nonlinear effects on the evolution of the instability, we also perform numerical simulations of the full lubrication equation using different seeds to generate the random sequence of amplitudes for the stochastic term (so that a realization corresponds to each seed), and average the resulting power spectra to obtain a representative spectrum to be compared with the one predicted by the LSA. As expected, we find a good agreement with LSA for early times. Interestingly, for late times we obtain that the wavenumber of the maximum of the spectra tends to approach the deterministic value, $q_m$, corresponding to the LSA without stochasticity. Since the LSA with stochasticity also tends to $q_m$, we can conclude that the typical lengths of the patterns in advanced stages of the instability with stochasticity seem to be close to the length of maximum growth rate of the linear deterministic modes. 

Therefore, encouraged by this result we also compare the LSA prediction with the experimental data from the instability of melted copper films on a silicon oxide substrate. These data correspond to the early stages, where the holes start to grow, as well as to the stages of drops formation, i.e. after having passed through the processes of film breakup and dewetting. A special feature of these data is that they come from different spatial regions of the laser spot, and thus received distinct illuminations. Thus, different times of a single evolution can be attributed to each region. These times were estimated here by fitting the LSA power spectrum to each experimental one with its corresponding value of $\beta$. As a result, we found that the early stages of this experiment evolved with a practically white noise in space, while a strong spatial correlation appeared in the spectra for late times. This shows that the explanation of experimental results in the nanometric scale requires the inclusion of some thermal noise in the modeling. In particular, correlated noise seems to be an important factor in the central regions of the laser spot, i.e. those with larger liquid lifetimes. We believe that our results justify further testing with more detailed experimental data.

\appendix
\section{Eigenvalues of the correlation function}
\label{app:eigenv}

Here, we calculate the eigenvalues of the Hilbert-Schmidt operator ${\cal Q}$ as defined by given by Eqs.~(\ref{eq:sigmak0}) and (\ref{eq:F}). By using the variable $v=\pi u/L$, the eigenvalues can be written as 
\begin{equation}
 \chi_k=\frac{A(\alpha,k)}{A(\alpha,0)}
 \label{eq:sigmak1}
\end{equation}
where 
\begin{equation}
 A(\alpha,k)= \int_0^\pi e^{-2\alpha (\sin v)^2 - 2 i k v}  dv,
 \label{eq:Aint}
\end{equation}
and $\alpha$ is given by Eq.~(\ref{eq:alpha}). In order to perform the above integral, we make the change of variables $2v = \theta+\frac{\pi}{2}$, which leads to the following expression, 
\[ \sin^2 v= \frac{1}{2} (1- \cos 2v)=\frac{1}{2} (1+ \sin \theta ).\]
This one allows us to write Eq.~(\ref{eq:Aint}) in terms of $\sin \theta$, as
\begin{equation}
A(\alpha,k)=\frac{1}{2}e^{-\alpha}e^{-\imath k \pi/2}\int_{-\frac{\pi}{2}}^{\frac{3\pi}{2}} 
e^{-\alpha\sin\theta}e^{-\imath k \theta} d\theta=\frac{(-\imath)^k}{2}e^{-\alpha}
\int_{-\frac{\pi}{2}}^{\frac{3\pi}{2}} e^{-\alpha\sin\theta}e^{-\imath k \theta} d\theta.
\label{eq:Ageneral}
\end{equation}
The above substitution is convenient in view of the the relation,
\begin{equation}
e^{\imath x \sin \theta}=\sum_{-\infty}^{\infty} e^{\imath m\theta} J_m(x),
\label{eq:generatrix}
\end{equation}
which becomes useful here upon defining $x=i\alpha$. Thus, we have
\begin{equation}
e^{-\alpha \sin \theta}=\sum_{-\infty}^{\infty} e^{\imath m\theta} J_m(\imath  \alpha),
\label{eq:generatrix1}
\end{equation}
where $J_m(\imath \alpha)$ is the Bessel function of order $m$. Now, we can also use the property 
\begin{equation}
J_m(\imath \alpha)=\imath^m I_m(\alpha),
\label{eq:JmIm}
\end{equation}
where $I_m(\alpha)$ is the modified Bessel function of order $m$. By replacing Eqs.~(\ref{eq:generatrix1}) and (\ref{eq:JmIm}) into Eq.~(\ref{eq:Ageneral}), we obtain
\begin{equation}
A(\alpha,k)=\frac{(-\imath)^k}{2}e^{-\alpha}\sum_{-\infty}^{\infty} \imath^m I_m(\alpha) 
\int_{-\frac{\pi}{2}}^{\frac{3\pi}{2}} e^{\imath (m-k)\theta} d\theta.
\end{equation}
Since, the above integral yields $2\pi \delta_{km}$, we finally have 
\begin{equation}
A(\alpha,k)=\pi e^{-\alpha}I_k(\alpha),
\end{equation}
so that the eigenvalue in Eq.(\ref{eq:sigmak1}) becomes
\begin{equation}
 \chi(q_k)=\chi_k=\frac{I_k(\alpha)}{I_0(\alpha)},
 \label{eq:sigmak2_app}
\end{equation}
which is the expression in Eq.~(\ref{eq:sigmak2}).

\bibliographystyle{unsrt}

\end{document}